\documentstyle[floats,prd,aps,epsf,eqsecnum,12pt]{revtex}

\makeatletter
\newbox\tempboxa
\newdimen\captionboxsubcount 
\def\capsize#1{\captionboxsubcount=#1pt}
\newdimen\captionboxsub
\captionboxsub=\hsize \advance\captionboxsub by -\captionboxsubcount
\advance\captionboxsub by -\captionboxsubcount
\long\def\@makecaption#1#2{
 \setbox\@tempboxa\hbox{#1 #2}
 \ifdim \wd\@tempboxa >\captionboxsub 
\rightskip=\captionboxsubcount \leftskip=\captionboxsubcount #1 #2 
\else \hbox to\hsize{\hfil\box\@tempboxa\hfil} 
 \fi}
\makeatother
\capsize{30}

\begin{document}

\begin{titlepage}

\begin{flushright}
\begin{minipage}{5cm}
\begin{flushleft}
\small
\baselineskip = 13pt
SU--4240--659 \\
UNITU--THEP--6/1997\\
hep-ph/9704358 \\
April, 1997
\end{flushleft}
\end{minipage}
\end{flushright}

\begin{center}
\Large\bf
Generalization of the Bound State Model
\end{center}

\vfil

\footnotesep = 12pt

\begin{center}
\large
Masayasu {\sc Harada}$^{(a)}$\footnote{
Electronic address : {\tt mharada@npac.syr.edu}
}, \quad
Francesco {\sc Sannino}$^{(a,b)}$\footnote{
Electronic address : {\tt sannino@npac.syr.edu}},
\\
Joseph {\sc Schechter}$^{(a)}$\footnote{
Electronic address : {\tt schechter@suhep.phy.syr.edu}}
\quad and \quad
Herbert {\sc Weigel}$^{(c)}$\footnote{
Electronic address : 
{\tt weigel@sunelc1.tphys.physik.uni-tuebingen.de}}
\end{center}

\begin{flushleft}
\it 
\qquad$^{(a)}$Department of Physics, Syracuse University, 
Syracuse, NY 13244-1130, USA
\\
\qquad$^{(b)}$%
Dipartimento di Scienze Fisiche \& Istituto
Nazionale di Fisica Nucleare \\
\qquad\qquad Mostra D'Oltremare Pad. 19,  80125 Napoli, Italy
\\
\qquad$^{(c)}$%
Institute for Theoretical Physics, 
T\"ubingen University \\
\qquad\qquad
Auf der Morgenstelle 14, D-72076 T\"ubingen, Germany
\end{flushleft}

\vfil

\begin{center}
\bf
Abstract
\end{center}

\begin{abstract}
\baselineskip = 17pt
In the bound state approach the heavy baryons are constructed by
binding, with any orbital angular momentum, the heavy meson multiplet
to the nucleon considered as a soliton in an effective 
meson theory.  
We point out that this picture misses an entire family of states,
labeled by a different angular momentum quantum number, which are
expected to exist according to the geometry of the three--body
constituent quark model (for $N_{\rm C}=3$).  
To solve this problem we propose that the bound state model be
generalized to include orbitally excited heavy mesons bound to the
nucleon.
In this approach the missing angular momentum is ``locked--up'' in the
excited heavy mesons.
In the simplest dynamical realization of the picture we give
conditions on a set of coupling constants for the binding of the
missing heavy baryons of arbitrary spin.
The simplifications made include working in the large $M$ limit,
neglecting nucleon recoil corrections, neglecting mass differences
among different heavy spin multiplets and also neglecting the
effects of light vector mesons.
\end{abstract}

\begin{flushleft}
\end{flushleft}

\end{titlepage}

\setcounter{footnote}{0}

\section{Introduction}

There has been recent interest in studying heavy baryons (those with
the quark structure $qqQ$) in the bound state
picture~\cite{Callan-Klebanov,Blaizot-Rho-Scoccola} together with
heavy quark spin symmetry~\cite{Eichten-Feinberg}.
In this picture the heavy baryon is treated~\cite{%
Guralnik-Luke-Manohar,Rho,%
Gupta-Momen-Schechter-Subbaraman,Schechter-Subbaraman,%
Schechter-Subbaraman-Vaidya-Weigel,Oh-Park} 
as a heavy spin multiplet of
mesons ($Q\bar{q}$) bound in the background field of the nucleon 
($qqq$), which in turn arises as a soliton configuration of light 
meson fields.

A nice feature of this approach is that it permits, in principle, an
exact expansion of the heavy baryon properties in simultaneous powers
of $1/M$ and $1/N_{\rm C}$.
In the simplest treatments, the light part of the chiral Lagrangian 
is made from only 
pion fields.
However it has been shown that the introduction of light vector
mesons~\cite{%
Gupta-Momen-Schechter-Subbaraman,Schechter-Subbaraman,%
Schechter-Subbaraman-Vaidya-Weigel}
substantially improves the accuracy of the model.
This is also true for the soliton treatment of the nucleon
itself~\cite{JJPSW,JPSSW,PW}.
Furthermore finite $M$ corrections as well as finite
$N_{\rm C}$ (nucleon recoil) corrections are also important.
This has been recently demonstrated for the hyperfine splitting
problem~\cite{HQSSW:1,HQSSW:2}.

Since the bound state--soliton approach is somewhat involved
it may be worthwhile to point out a couple of its advantages.
In the first place, it is based on an effective chiral Lagrangian
containing physical parameters which are in principle subject to
direct experimental test.
Secondly, the bound state approach models a characteristic feature of
a confining theory.
When the bound system is suitably ``stretched'' it does not separate
into colored objects but into physical color singlet states.

Here we shall investigate the spectrum of excited states in the bound
state--soliton framework.
Some aspects of this problem have already been 
treated~\cite{CW,Schechter-Subbaraman,Oh-Park,HQSSW:1,HQSSW:2}.
We will deal with an aspect which does not seem to have been
previously discussed.
This emerges when one compares the excited heavy baryon spectrum with
that expected in the constituent quark model (CQM)~\cite{CI}.
We do not have in mind specific dynamical treatments of the CQM 
but rather just its general geometric structure.
Namely we shall just refer to the counting of states which follows
from considering the baryon as a three body system obeying
Fermi--Dirac statistics.  We shall restrict our attention to the 
physical states for $N_{\rm C}=3$.
In this framework the CQM counting of the heavy excited baryon
multiplets has been recently discussed~\cite{Koerner}.
At the level of two light flavors there are expected to be seven
negative parity first excited $\Lambda$--type heavy baryons and seven
negative parity first excited $\Sigma$--type heavy baryons.
On the other hand a similar counting~\cite{%
Schechter-Subbaraman,HQSSW:2} in the bound state treatments 
mentioned above yields only two of the $\Lambda$--type and five of
the $\Sigma$--type.  Thus there are seven missing first excited states.
One thought is that these missing states should be unbound and thus
represent new dynamical information with respect to the simple
geometrical picture. There is certainly not enough data for the 
charmed baryons to decide this issue.  However for the strange 
baryons there are ten established particles for these fourteen 
states. Hence it is reasonable to believe that these states exist 
for the heavy baryons too. In the CQM one may have two different 
sources of orbital angular momentum excitation; for example the 
relative angular momentum of the two light quarks, $L_I$ and the 
angular momentum, $L_E$ of the diquark system with respect to the
heavy quark.  The parity of the heavy baryon is given by 
$P=\left(-1\right)^{L_I+L_E}$. However, in the bound state models 
considered up to now there is only room for one relative angular 
momentum, $r$ associated with the wave function of the heavy meson 
with respect to the soliton. The parity is given by 
$P=\left(-1\right)^r$.  Both models agree on the counting of the 
``ground'' states ($L_I=L_E=r=0$).  Also the counting of the states 
with ($L_I=0$, $L_E=1$) agrees with those of $r=1$ in the bound state 
model.  However, the bound state model has no analog of the ($L_I=1$, 
$L_E=0$) states and, in general, no analog of the higher $L_I\neq0$ 
states either.

It is clear that we must find a way of incorporating a new angular
momentum quantum number in the bound state picture. One might imagine 
a number of different ways to accomplish this goal.  Here we will 
investigate a method which approximates a three body problem by an 
effective two body problem. Specifically we will consider binding 
excited heavy mesons with orbital angular momentum $\ell$ to the
soliton.
The excited heavy mesons may be interpreted as bound states of 
the original heavy meson and a surrounding light meson cloud.
Then the baryon parity comes out to be $\left(-1\right)^{r+\ell}$.
This suggests a correspondence (but not an identity) $r
\leftrightarrow L_E$, $\ell \leftrightarrow L_I$ and additional new
states.
An interesting conceptual point of the model is that it displays a
correspondence between the excited heavy mesons and the excited heavy
baryons.

Almost immediately one sees that the model is considerably more
complicated than the previous one in which the single heavy field
multiplet $H$ is bound to the soliton.
Now, for each value of $\ell\neq0$, there will be two different higher 
spin heavy multiplets which can contribute.
In fact there is also a mixing between multiplets with different
$\ell$, which is therefore not actually a good quantum number for the
model (unless the mixing is neglected).

Thus we will make a number of approximations which seem reasonable for
an initial analysis.
For one thing we shall neglect the light vector mesons even though we
know they may be important.
We shall also neglect the possible effects of higher spin light
mesons, which one might otherwise consider natural when higher spin
heavy mesons are being included.
Since there is a proliferation of interaction terms among the light
and heavy mesons we shall limit ourselves to those with the minimum
number of derivatives.
Finally, $1/M$ and nucleon recoil corrections will be neglected.
The resulting model is the analog of the initial one used previously.
Even though the true picture is likely to be more involved than our
simplified model, we feel that the general scheme presented here will
provide a useful guide for further work.

We would like to stress that this bound state model goes beyond the
kinematical enumeration of states and contains dynamical information.
Specifically, the question of which states are bound depends on the
magnitudes and signs of the coupling constants.
There is a choice of coupling constants yielding a natural pattern of
bound states which includes the missing ones.
It turns out that it is easier to obtain the precise missing state
pattern for the $\Lambda$--type heavy particles.
Generally, there seem to be more than just the missing $\Sigma$--type
heavy baryons present.
However we show that the collective quantization, which is anyway
required in the bound state approach, leads to a splitting which may
favor the missing heavy spin multiplets.

This paper is organized in the following way.
Section~\ref{sec:preliminaries} starts with a review of the CQM
geometrical counting of excited heavy baryon multiplets.
It continues with a quick summary of the treatment of heavy baryons in
the existing bound state models.
The comparison of the mass spectrum in the two different approaches
reveals that there is a large family of ``missing'' excited states.
This is discussed in general terms in section~\ref{sec: missing} where
a proposal for solving the problem by considering the binding of heavy
excited mesons to the Skyrmion is made.
A correspondence between the angular momentum variables of the CQM and
of the new model is set up.
A detailed treatment of the proposed model for the case of the first
excited heavy baryons is given in section~\ref{sec:4}.
This includes discussion of the heavy meson bound state wave function,
the classical potential energy as well as the energy corrections due
to quantization of the collective variables of the model.
It is pointed out that there is a possible way of choosing the
coupling constants so as to bind all the missing states.
The generalization to the excited heavy baryon states of arbitrary
spin is given in section~\ref{sec:5}.
This section also contains some new material on the interactions of
the heavy meson multiplets with light chiral fields.
Section~\ref{sec:6} contains a discussion of the present status of the
model introduced here.
Finally, some details of the calculations are given in
Appendices~\ref{app:a} and \ref{app:b}.

\section{Some Preliminaries}
\label{sec:preliminaries}

In this section, for the reader's convenience, we will briefly discuss
which heavy baryon states are predicted by the CQM as well as some 
relevant material needed for the bound state approach to the heavy
baryon states. 

It is generally agreed that the geometrical structure of the CQM 
provides a reasonable guide for, at least, counting and labeling 
the physical strong interaction ground states. When radial excitations
or dynamical aspects are considered the model predictions are 
presumably less reliable. In the CQM the heavy baryons consist of two 
light quarks ($q$) and a heavy quark ($Q$) in a color singlet state.
Since the color singlet states are antisymmetric on interchange of the
color labels of any two quarks, the overall wave function must,
according to Fermi-Dirac statistics, be fully symmetric on interchange
of flavor, spin and spatial indices.
Here we will consider the case of two light flavors.
For counting the states we may choose coordinates~\cite{Koerner}
so that the total angular momentum of the heavy baryon,
$\mbox{\boldmath$J$}$ is decomposed as
\begin{equation}
\mbox{\boldmath$J$} = \mbox{\boldmath$L$}_I + \mbox{\boldmath$L$}_E
+ \mbox{\boldmath$S$} + \mbox{\boldmath$S$}_H \ ,
\label{baryon spin:CQM}
\end{equation}
where $\mbox{\boldmath$L$}_I$ represents the relative orbital angular
momentum of the two light quarks, $\mbox{\boldmath$L$}_E$ the orbital
angular momentum of the light diquark center of mass with respect to
the heavy quark, $\mbox{\boldmath$S$}$ the total spin of the diquarks
and $\mbox{\boldmath$S$}_H$ the spin of the heavy quark. In the 
``heavy'' 
limit where the heavy quark becomes infinitely
massive $\mbox{\boldmath$S$}_H$ 
completely decouples. The parity of the heavy baryon is given by
\begin{equation}
P_B = \left( -1 \right)^{L_I + L_E} \ .
\label{baryon parity:CQM}
\end{equation}
Since we are treating only the light degrees of freedom as identical
particles it is only necessary to symmetrize the diquark product wave
function with respect to the $\mbox{\boldmath$L$}_I$,
$\mbox{\boldmath$S$}$ and isospin $\mbox{\boldmath$I$}$ labels.
Note that the diquark isospin $\mbox{\boldmath$I$}$ equals 
the baryon isospin. 
There are four possible ways to build an overall wave function
symmetric with respect to these three labels:
\begin{eqnarray}
&&\mbox{a)} \quad I=0 \ , \quad S=0 \ , \quad L_I = \mbox{even} \ ,
\nonumber\\
&&\mbox{b)} \quad I=1 \ , \quad S=1 \ , \quad L_I = \mbox{even} \ ,
\nonumber\\
&&\mbox{c)} \quad I=0 \ , \quad S=1 \ , \quad L_I = \mbox{odd} \ ,
\nonumber\\
&&\mbox{d)} \quad I=1 \ , \quad S=0 \ , \quad L_I = \mbox{odd} \ .
\label{condition:CQM}
\end{eqnarray}
There is no kinematic restriction on $L_E$.\footnote{%
We are adopting a convention where bold--faced angular momentum
quantities are vectors and the regular quantities stand for their
eigenvalues.}

Let us count the possible baryon states. The 
$L_I=L_E=0$ heavy baryon ground state consists of $\Lambda_Q$
($J^P=\frac{1}{2}^+$) from a) and the heavy spin multiplet
$\left\{\Sigma_Q\left(\frac{1}{2}^+\right) \ , \ 
\Sigma_Q\left(\frac{3}{2}^+\right) \right\}$ from b).
It is especially interesting to consider the first orbitally excited
states.  These all have negative parity with either ($L_E=1$, $L_I=0$) or
($L_E=0$, $L_I=1$).
For $L_E=1$, a) provides the heavy spin multiplet 
$\left\{\Lambda_Q\left(\frac{1}{2}^-\right)\ , 
\ \Lambda_Q\left(\frac{3}{2}^-\right)\right\}$ and b) provides
$\Sigma_Q\left(\frac{1}{2}^-\right)$,
$\left\{\Sigma_Q\left(\frac{1}{2}^-\right)\ ,
\ \Sigma_Q\left(\frac{3}{2}^-\right)\right\}$,
$\left\{\Sigma_Q\left(\frac{3}{2}^-\right)\ ,
\ \Sigma_Q\left(\frac{5}{2}^-\right)\right\}$.
For $L_I=1$ c) provides $\Lambda_Q\left(\frac{1}{2}^-\right)$,
$\left\{\Lambda_Q\left(\frac{1}{2}^-\right)\ ,\
\Lambda_Q\left(\frac{3}{2}^-\right)\right\}$,
$\left\{\Lambda_Q\left(\frac{3}{2}^-\right)\ ,\
\Lambda_Q\left(\frac{5}{2}^-\right)\right\}$,
while d) provides
$\left\{\Sigma_Q\left(\frac{1}{2}^-\right)\ ,
\ \Sigma_Q\left(\frac{3}{2}^-\right)\right\}$.
Altogether there are fourteen different isotopic spin multiplets at
the first excited level.
The higher excited levels can be easily enumerated in the same way.
For convenient reference these are listed in Table~\ref{table:1}.
\begin{table}[htbp]
\begin{center}
\begin{tabular}{c|cc}
\multicolumn{1}{c}{\ } & $L_E=0$ & $L_E=1$ \\
\hline
 $L_I=0$ 
& \multicolumn{1}{c}{$\begin{array}{c}
\Lambda_Q\left(\frac{1}{2}^+\right) \\
\left\{\Sigma_Q\left(\frac{1}{2}^+\right) \,,\,
\Sigma_Q\left(\frac{3}{2}^+\right)\right\}
\end{array}$ }
& \multicolumn{1}{c}{$\begin{array}{c}
\left\{\Lambda_Q\left(\frac{1}{2}^-\right) \,,\,
\Lambda_Q\left(\frac{3}{2}^-\right)\right\} \\
\Sigma_Q\left(\frac{1}{2}^-\right) \\
\left\{\Sigma_Q\left(\frac{1}{2}^-\right) \,,\,
\Sigma_Q\left(\frac{3}{2}^-\right)\right\} \\
\left\{\Sigma_Q\left(\frac{3}{2}^-\right) \,,\,
\Sigma_Q\left(\frac{5}{2}^-\right)\right\}
\end{array}$ }
\\
\hline
 $L_I=1$
& \multicolumn{1}{c}{$\begin{array}{c}
\Lambda_Q\left(\frac{1}{2}^-\right) \\
\left\{\Lambda_Q\left(\frac{1}{2}^-\right) \,,\,
\Lambda_Q\left(\frac{3}{2}^-\right)\right\} \\
\left\{\Lambda_Q\left(\frac{3}{2}^-\right) \,,\,
\Lambda_Q\left(\frac{5}{2}^-\right)\right\} \\
\left\{\Sigma_Q\left(\frac{1}{2}^-\right) \,,\,
\Sigma_Q\left(\frac{3}{2}^-\right)\right\}
\end{array}$ }
& $\cdots$ 
\\
\hline
 $\vdots$ & & 
\\
\hline
 $L_I=2n-1$
& \multicolumn{1}{c}{$\begin{array}{c}
\left\{\Lambda_Q\left(\left(2n-\frac{5}{2}\right)^-\right) \,,\,
\Lambda_Q\left(\left(2n-\frac{3}{2}\right)^-\right)\right\} \\
\left\{\Lambda_Q\left(\left(2n-\frac{3}{2}\right)^-\right) \,,\,
\Lambda_Q\left(\left(2n-\frac{1}{2}\right)^-\right)\right\} \\
\left\{\Lambda_Q\left(\left(2n-\frac{1}{2}\right)^-\right) \,,\,
\Lambda_Q\left(\left(2n+\frac{1}{2}\right)^-\right)\right\} \\
\left\{\Sigma_Q\left(\left(2n-\frac{3}{2}\right)^-\right) \,,\,
\Sigma_Q\left(\left(2n-\frac{1}{2}\right)^-\right)\right\}
\end{array}$ }
& $\cdots$ 
\\
\hline
 $L_I=2n$
& \multicolumn{1}{c}{$\begin{array}{c}
\left\{\Lambda_Q\left(\left(2n-\frac{1}{2}\right)^+\right) \,,\,
\Lambda_Q\left(\left(2n+\frac{1}{2}\right)^+\right)\right\} \\
\left\{\Sigma_Q\left(\left(2n-\frac{3}{2}\right)^+\right) \,,\,
\Sigma_Q\left(\left(2n-\frac{1}{2}\right)^+\right)\right\} \\
\left\{\Sigma_Q\left(\left(2n-\frac{1}{2}\right)^+\right) \,,\,
\Sigma_Q\left(\left(2n+\frac{1}{2}\right)^+\right)\right\} \\
\left\{\Sigma_Q\left(\left(2n+\frac{1}{2}\right)^+\right) \,,\,
\Sigma_Q\left(\left(2n+\frac{3}{2}\right)^+\right)\right\} 
\end{array}$ }
& $\cdots$ 
\\
\hline
 $\vdots$ & & 
\\
\end{tabular}
\end{center}
\caption[]{
Examples of the heavy baryon multiplets predicted by the CQM.
}
\label{table:1}
\end{table}

It is natural to wonder whether all of these states should actually
exist experimentally. This is clearly a premature question for the $c$
and $b$ baryons.  However an indication for the first excited states 
can be gotten from the
ordinary hyperons (or $s$ baryons).  In this case there are six well
established candidates~\cite{PDG} for the $\Lambda$'s 
[$\Lambda(1405)$, $\Lambda(1520)$, $\Lambda(1670)$, $\Lambda(1690)$,
$\Lambda(1800)$ and $\Lambda(1830)$]; only one $\frac{3}{2}^-$ state
has not yet been observed.  For the $\Sigma$'s there are four well
established candidates [$\Sigma(1670)$, $\Sigma(1750)$, $\Sigma(1775)$
and $\Sigma(1940)$]; two $\frac{1}{2}^-$ states and one
$\frac{3}{2}^-$ state have not yet been observed.  
Thus it seems plausible to expect that all fourteen of the first 
excited negative parity heavy baryons do indeed exist.
We might also expect higher excited states to exist.

What is the situation in the bound state approach?
To study this we shall briefly summarize the usual approach~\cite{%
Schechter-Subbaraman,Oh-Park,HQSSW:2}
to the excited heavy baryons in the bound state picture.
In this model the heavy baryon is considered to be a heavy meson
bound, via its interactions with the light mesons, to a nucleon
treated as a Skyrme soliton.
The model is based on a chiral Lagrangian with two parts, 
${\cal L} = {\cal L}_{\rm light} + {\cal L}_{\rm heavy}$.
The light part involves the chiral field
$U=\xi^2=\exp\left(2i\phi/F_\pi\right)$,
where $\phi$ is the $2\times2$ matrix of standard pion fields. 
Relevant vector and pseudovector combinations are  
\begin{equation}
v_\mu \ , \ p_\mu = 
\frac{i}{2} \left( \xi \partial_\mu \xi^{\dag}
\pm \xi^{\dag} \partial_\mu \xi \right)
\ .
\end{equation}
In addition light vector mesons are included in a $2\times2$ matrix
field $\rho_\mu$, which describes both the rho and omega particles.
The light Lagrangian has a classical soliton solution of the form
\begin{eqnarray}
&&
\xi_{\rm c} (\mbox{\boldmath$x$}) = 
\exp \left[ \frac{i}{2} \hat{\mbox{\boldmath$x$}}\cdot
\mbox{\boldmath$\tau$} \, F(|\mbox{\boldmath$x$}|) \right]
\ ,
\nonumber\\
&&
\rho^a_{i{\rm c}} = 
\frac{1}{\sqrt{2}\tilde{g}\vert\mbox{\boldmath$x$}\vert}
\epsilon_{ika} \hat{x}_k G(\vert\mbox{\boldmath$x$}\vert)
\ ,
\nonumber\\
&&
\omega_{0{\rm c}} = \omega(\vert\mbox{\boldmath$x$}\vert)
\ ,
\nonumber\\
&&
\rho^a_{0{\rm c}} = \omega_{i{\rm c}} = 0 \ ,
\label{xi classical}
\end{eqnarray}
where
$\rho_{\mu{\rm c}} = \frac{1}{2}
\left( \omega_{\mu{\rm c}} + \tau^a \rho^a_{\mu{\rm c}} \right)$
and $\tilde{g}$ is a coupling constant.
The appropriate boundary conditions are
\begin{eqnarray}
&&
F(0) = -\pi \ , \quad G(0) = 2 \ , \quad \omega^\prime(0) = 0 \ ,
\nonumber\\
&&
F(\infty) = G(\infty) = \omega(\infty) = 0 \ , 
\end{eqnarray}
which correspond to unit baryon number.

The heavy Lagrangian will be constructed, to insure heavy spin
symmetry, from the fluctuation field $H$ describing the heavy
pseudoscalar and vector mesons.
It takes the form~\cite{Schechter-Subbaraman:2}
\begin{equation}
{\cal L}_{\rm heavy}/M
= i V_\mu \mbox{Tr}\, \left[ H D_\mu \bar{H} \right]
+ i d \, \mbox{Tr}\, \left[ H \gamma_\mu \gamma_5 p_\mu \bar{H}
\right] 
+ \frac{ic}{m_V} \mbox{Tr} \, \left[
H \gamma_\mu \gamma_\nu F_{\mu\nu}(\rho) \bar{H} \right]
\ ,
\label{Lag for H}
\end{equation}
where $D_\mu \equiv \partial_\mu - i \alpha \tilde{g} \rho_\mu
- i (1-\alpha) v_\mu$, $V_\mu$ is the four
velocity of the heavy meson and $F_{\mu\nu}(\rho) = \partial_\mu
\rho_\nu - \partial_\nu \rho_\mu - i \tilde{g} 
\left[ \rho_\mu \,,\, \rho_\nu \right]$.
Furthermore, $m_V$ is the light vector meson mass while
$d\simeq0.53$ and $c\simeq1.6$ are respectively the heavy
meson--pion and magnetic type heavy meson--light vector meson coupling
constants; $\alpha$ is a coupling constant whose value has not yet
been firmly 
established.  Previous work has shown~\cite{%
Gupta-Momen-Schechter-Subbaraman,Schechter-Subbaraman,%
Schechter-Subbaraman-Vaidya-Weigel,HQSSW:2}
that a quantitatively more accurate description of the heavy baryons
is obtained when light vector mesons are included in ${\cal L}$.

The wave function for the heavy meson bound to the background Skyrmion
field (\ref{xi classical}) is conveniently presented in the rest
frame, $\mbox{\boldmath$V$}=0$.
In this frame
\begin{equation}
\bar{H}_{\rm c} \rightarrow
\left( \begin{array}{cc}
0 & 0 \\ \bar{h}_{lh}^a & 0 
\end{array}\right) 
\ ,
\label{H: matrix}
\end{equation}
with $a$, $l$, $h$ representing respectively the isospin, light spin
and heavy spin bivalent indices.
The calculation simplifies if we deal with a radial wave function
obtained after removing the factor
$\hat{\mbox{\boldmath$x$}}\cdot\mbox{\boldmath$\tau$}$:
\begin{equation}
\bar{h}_{lh}^a = \frac{u(|\mbox{\boldmath$x$}|)}{\sqrt{M}}
\left(
  \hat{\mbox{\boldmath$x$}}\cdot\mbox{\boldmath$\tau$}
\right)_{ad}
\psi_{dl} \chi_h
\ ,
\label{h: classical}
\label{Wf:h bar}
\end{equation}
where $u(|\mbox{\boldmath$x$}|)$ is a radial wave function, assumed 
to be very sharply peaked near $|\mbox{\boldmath$x$}|=0$ for large $M$.
The heavy spinor $\chi_h$ is trivially factored out in this expression
as a manifestation of the heavy quark symmetry.
We perform a partial wave analysis of the generalized ``angular'' wave
function $\psi_{dl}$:
\begin{equation}
\psi_{dl}\left(g,g_3;r,k\right) = 
\sum_{r_3,k_3} C_{r_3,k_3;g_3}^{r,k;g}
Y^{r_3}_r \xi_{dl}(k,k_3) \ .
\label{def:wave function}
\end{equation}
Here $Y^{r_3}_r$ stands for the standard spherical harmonic
representing orbital angular momentum $r$ while $C$ denotes the
ordinary Clebsch--Gordan coefficients.
$\xi_{dl}(k,k_3)$ represents a wave function in which the ``light
spin'' and isospin (referring to the ``light cloud'' component 
of the heavy meson) are added vectorially to give
\begin{equation}
\mbox{\boldmath$K$} = \mbox{\boldmath$I$}_{\rm light}
+ \mbox{\boldmath$S$}_{\rm light} \ ,
\label{def: k}
\end{equation}
with eigenvalues $\mbox{\boldmath$K$}^2=k(k+1)$.
The total light ``grand spin''
\begin{equation}
\mbox{\boldmath$g$} = \mbox{\boldmath$r$} +
\mbox{\boldmath$K$} 
\label{def: g1}
\end{equation}
is a significant quantity in the heavy limit.

Substituting the wave--function (\ref{Wf:h bar}) into 
$\int\,d^3x\,{\cal L}_{\rm heavy}$ given in Eq.~(\ref{Lag for H})
yields the potential operator
\begin{eqnarray}
V &=& \int d \Omega \ \psi^{\ast} 
\left\{
  \mbox{\boldmath$\sigma$}\cdot\mbox{\boldmath$\tau$} 
  \Delta_1 + 1\, \Delta_2
\right\} \psi
\nonumber\\
&=&
\int \, d\Omega\ \psi^{\ast}
\left\{
  4 \Delta_1 \mbox{\boldmath$S$}_{\rm light} \cdot
  \mbox{\boldmath$I$}_{\rm light} 
  + 1\, \Delta_2
\right\}
\psi
\nonumber\\
&=&
2 \Delta_1 \, \left[ k(k+1) - \frac{3}{2} \right] + \Delta_2
\ ,
\label{potential: H}
\end{eqnarray}
where $\int\,d\Omega$ is the solid angle integration and 
Eq.~(\ref{def: k}) was used in the last step.
In addition
\begin{eqnarray}
\Delta_1 &=& \frac{1}{2} d\, F'(0) - \frac{c}{m_V\tilde{g}}
G^{\prime\prime}(0) \ ,
\nonumber\\
\Delta_2 &=& - \frac{\alpha\tilde{g}}{\sqrt{2}} \omega(0)
\ .
\end{eqnarray}
The $\Delta_2$ term is relatively small~\cite{%
Schechter-Subbaraman,Schechter-Subbaraman-Vaidya-Weigel,HQSSW:2} 
and will be neglected.
Both terms in $\Delta_1$ are positive with the second one (due to
light vectors) slightly larger.
There are just the two possibilities $k=0$ and $k=1$.
It is seen that the $k=0$ states, for any orbital angular momentum
$r$, will be bound with binding energy 
$3\Delta_1$.
The $k=1$ states are unbound in this limit.
The parity of the bound state wave function is 
\begin{equation}
P_B = \left( -1 \right)^r \ ,
\label{baryon parity}
\end{equation}
which emerges as a product of $\left(-1\right)^r$ for $Y^{r_3}_r$ in 
Eq.~(\ref{def:wave function}), $-1$ for the 
$\hat{\mbox{\boldmath$x$}}\cdot\mbox{\boldmath$\tau$}$ factor in
Eq.~(\ref{Wf:h bar}) and $-1$ due to the fact that 
the mesons bound to the soliton have negative parity.

The states of definite angular momentum and isospin are generated, 
in the soliton approach, after collective quantization.
The collective angle--type coordinate $A(t)$ is introduced~\cite{Ad83}
as 
\begin{eqnarray}
\xi(\mbox{\boldmath$x$},t) 
&=&
A(t) \xi_{\rm c} (\mbox{\boldmath$x$}) A^{\dag}(t)
\ ,
\nonumber\\
\mbox{\boldmath$\tau$}\cdot\mbox{\boldmath$\rho$} 
\left( \mbox{\boldmath$x$}\,,\,t\right)
&=&
A(t) \mbox{\boldmath$\tau$}\cdot\mbox{\boldmath$\rho$}_{\rm c}
\left(\mbox{\boldmath$x$}\right) A^{-1}(t) \ ,
\nonumber\\
\bar{H}(\mbox{\boldmath$x$},t) 
&=&
A(t) \bar{H}_{\rm c} (\mbox{\boldmath$x$}) 
\ ,
\label{def: collective}
\end{eqnarray}
where $\xi_{\rm c}$ and $\mbox{\boldmath$\rho$}_{\rm c}$
are defined in Eq.~(\ref{xi classical}) and
$\bar{H}_{\rm c}$ in Eqs.~(\ref{H: matrix}) and
(\ref{h: classical}). For our purposes the important variable is the
``angular--velocity'' $\mbox{\boldmath$\Omega$}$ defined by
\begin{equation}
A^{\dag} \dot{A} = \frac{i}{2} 
\mbox{\boldmath$\tau$} \cdot \mbox{\boldmath$\Omega$}
\ ,
\label{def: omega}
\end{equation}
which measures the time dependence of the collective coordinates
$A(t)$. 
It should furthermore be mentioned that, due to the collective 
rotation, the vector meson field components which vanish classically 
($\rho^a_0$ and $\omega_i$) get induced.
For each bound state solution $\bar{H}_{\rm c}$, there will be a tower
of states characterized by a soliton angular momentum 
$\mbox{\boldmath$J$}^{\rm sol}$ and the total isospin 
$\mbox{\boldmath$I$}$ satisfying $I=J^{\rm sol}$.
The soliton angular 
momentum is computed from this collective Lagrangian as
\begin{equation}
\mbox{\boldmath $J$}^{\rm sol} = 
\frac{\partial L_{\rm coll}}{\partial \mbox{\boldmath $\Omega$}}
\ ,
\label{Jsol}
\end{equation}
while the total baryon angular momentum is the sum
\begin{equation}
\mbox{\boldmath$J$} =
\mbox{\boldmath$g$} + \mbox{\boldmath$J$}^{\rm sol} + 
\mbox{\boldmath$S$}_{\rm heavy}
\ ,
\label{baryon spin}
\end{equation}
where $\mbox{\boldmath$S$}_{\rm heavy}$ is the spin of the heavy quark
within the heavy meson.

Now we can list the bound states of this model.
First consider the $r=0$ states.
According to Eq.~(\ref{baryon parity}), they have positive parity.
Since Eq.~(\ref{potential: H}) shows that $k=0$ for binding,
Eq.~(\ref{def: g1}) tells us that the light ``grand spin'' $g=0$.
Equation~(\ref{baryon spin}) indicates (noting $I=J^{\rm sol}$) that
there will be a  $\Lambda_Q\left(\frac{1}{2}^+\right)$ state as well
as a $\left\{\Sigma_Q\left(\frac{1}{2}^+\right),\ 
\Sigma_Q\left(\frac{3}{2}^+\right)\right\}$
heavy spin multiplet.
Actually the model also predicts a whole tower of states with 
increasing isospin. Next there will be an $I=2$ 
heavy spin multiplet with spins and parity $\frac{3}{2}^+$ and 
$\frac{5}{2}^+$, and so forth. Clearly the isospin zero and one 
states correspond exactly to the $L_I=L_E=0$ ground states of the 
constituent quark model. The isotopic spin two states would also be 
present if we were to consider the ground state heavy baryons in 
a constituent quark model with number of colors, $N_{\rm C}=5$.
This is consistent with the picture~\cite{Ad83}
of the Skyrme model as a
description of the large $N_{\rm C}$ limit.

Next, consider the $r=1$ states.
These all have negative parity and (since the bound states have $k=0$)
light grand spin, $g=1$.
The $J^{\rm sol}=I=0$ choice yields a heavy multiplet
$\left\{\Lambda_Q\left(\frac{1}{2}^-\right)\ ,
\ \Lambda_Q\left(\frac{3}{2}^-\right)\right\}$
while the $J^{\rm sol}=I=1$ choice yields the three heavy multiplets
$\left\{\Sigma_Q\left(\frac{1}{2}^-\right)\right\}$,
$\left\{\Sigma_Q\left(\frac{1}{2}^-\right)\ , 
\Sigma_Q\left(\frac{3}{2}^-\right)\right\}$
and
$\left\{\Sigma_Q\left(\frac{3}{2}^-\right)\ , 
\Sigma_Q\left(\frac{5}{2}^-\right)\right\}$. These three multiplets
are associated with the intermediate sums
$|\mbox{\boldmath $g$}+\mbox{\boldmath $J$}^{\rm sol}|=0,1,2$,
respectively. It is evident that the seven states obtained have the 
same quantum numbers as the seven constituent quark states with 
$L_I=0$ and $L_E=1$.  Proceeding in the same way, it is easy to see 
that the bound states with general $r$ agree with those states in 
the constituent quark model which have $L_I=0$ and $L_E=r$.
This may be understood by rewriting Eqs.~(\ref{baryon spin}) and
(\ref{def: g1}) as
\begin{equation}
\mbox{\boldmath$J$} =
\mbox{\boldmath$r$} + \mbox{\boldmath$J$}^{\rm sol} +
\mbox{\boldmath$S$}_{\rm heavy}
\ ,
\end{equation}
where $k=0$ for the bound states was used.
Comparing this with the $L_I=0$ limit of the constituent quark model
relation (\ref{baryon spin:CQM}) shows that there seems to be a
correspondence
\begin{eqnarray}
\mbox{\boldmath$S$}_{\rm heavy}
&\leftrightarrow&
\mbox{\boldmath$S$}_H
\ ,
\nonumber\\
\mbox{\boldmath$r$}
&\leftrightarrow&
\mbox{\boldmath$L$}_E
\ ,
\nonumber\\
\mbox{\boldmath$J$}^{\rm sol}
&\leftrightarrow&
\mbox{\boldmath$S$}
\ .
\label{correspondence:1}
\end{eqnarray}
This correspondence is reinforced when we notice that 
$I=J^{\rm sol}$ in the bound state model and,
for the relevant cases a) and b) in Eq.~(\ref{condition:CQM}) of the
constituent quark model, $I=S$ also.
We stress that Eq.~(\ref{correspondence:1}) is a correspondence rather
than an exact identification of the same dynamical variables in
different models. It should be remarked that in the exact heavy and
large 
$N_c$ limits the heavy baryons for all values of $r=g$ will have the 
same mass. When finite $1/M$ corrections are taken into account, there 
will always be, in addition to other things, a ``centrifugal term'' in 
the effective potential of the form 
${g(g+1)}/({2M\vert\mbox{\boldmath$x$}\vert^2})$,
which makes the states with larger values of $g$, heavier.
It should also be remarked that the above described ordering of 
heavy baryon states in the bound state approach applies only
to the heavy limit, where $\mbox{\boldmath $S$}_{\rm heavy}$ 
decouples. For finite heavy quark masses, multiplets are 
characterized by the total grand spin 
$\mbox{\boldmath $g$}+
\mbox{\boldmath $S$}_{\rm heavy}$. Then states like
$\Lambda_Q\left(\frac{1}{2}^-\right)$ and 
$\Lambda_Q\left(\frac{3}{2}^-\right)$ no longer constitute 
a degenerate multiplet.

\section{The missing states}
\label{sec: missing}

It is clear that the bound state model discussed above 
contains only half
of the fourteen negative parity, first excited states predicted by the
CQM. The states with $L_I\neq0$ are all missing.
Since the enumeration of states in the CQM was
purely kinematical one might at first think that the bound state
model (noting that the dynamical condition $k=0$ was used) is
providing a welcome constraint on the large number of expected states.
However, experiment indicates that this is not likely to be the case.
As pointed out in the last section, there are at present good
experimental candidates for ten out of the fourteen negative parity,
first excited ordinary hyperons.
Thus the missing excited states appear to be a serious problem for the
bound state model.

\begin{figure}[htbp]
\begin{center}
\ \epsfbox{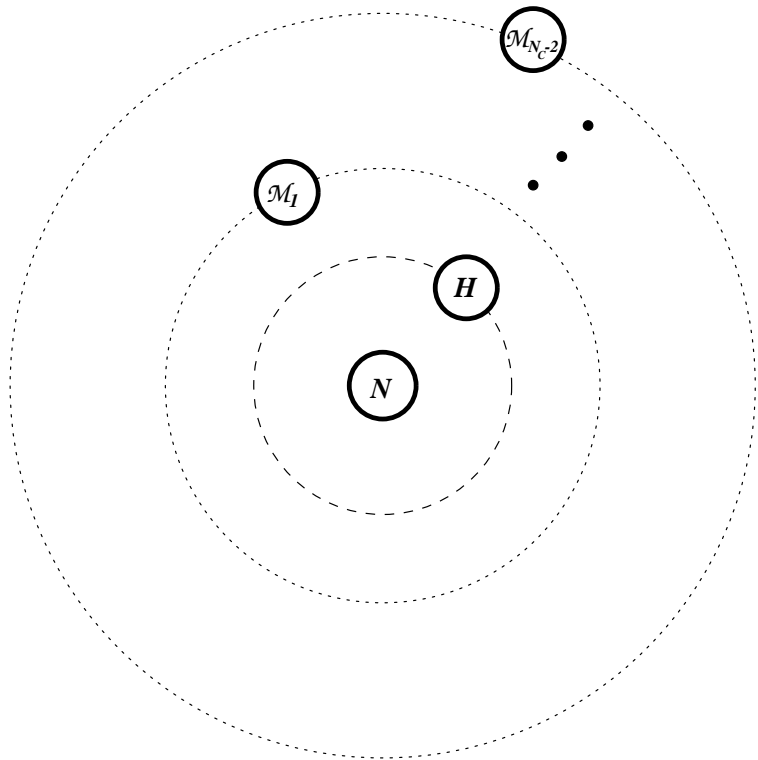}
\end{center}
\caption[]{Schematic planetary picture for large $N_{\rm C}$ excited
heavy baryons in the bound state approach.}
\label{fig:1}
\end{figure}

\begin{figure}[htbp]
\begin{center}
\ \epsfbox{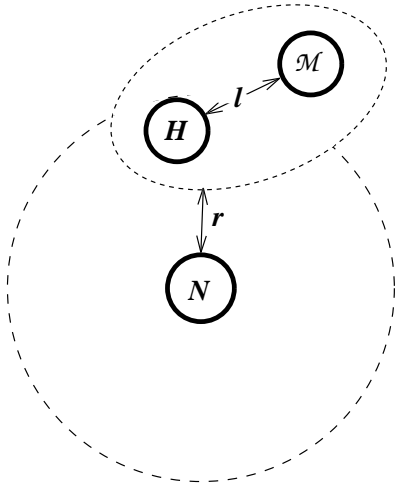}
\end{center}
\caption[]{Schematic picture of the ``two body'' approximation for the
$N_{\rm C}=3$ excited heavy baryons.}
\label{fig:2}
\end{figure}

The goal of the present paper is to find a suitable extension of the
bound state model which gives the same spectrum as the CQM.
Reference to Eq.~(\ref{baryon spin:CQM}) suggests that we introduce a
new degree of freedom which is related in some way to the light
diquark relative angular momentum $\mbox{\boldmath$L$}_I$.
To gain some perspective, and because we are working in a Skyrme model
overall framework, it is worthwhile to consider the heavy baryons in a
hypothetical world with $N_{\rm C}$ quark colors. In such a case 
there would be $N_{\rm C}-1$ relative angular momentum variables and 
we would require $N_{\rm C}-2$ additional degrees of freedom.
Very schematically we might imagine, as in Fig.~\ref{fig:1},
one heavy meson $H$ and $N_{\rm C}-2$ light mesons ${\cal M}_i$ orbiting
around the nucleon.
One might imagine a number of different schemes for treating the
inevitably complicated bound state dynamics of such a system.
Even in the $N_{\rm C}=3$ case it is much simpler if we can manage to
reduce the three body problem to an effective two body problem.
This can be achieved, as schematically indicated in Fig.~\ref{fig:2},
if we link the two ``orbiting'' mesons together in a state which
carries internal angular momentum.
The ``linked mesons'' will be described mathematically by a single 
excited heavy meson multiplet field. One may alternatively 
consider these ``linked mesons'' as bare heavy mesons surrounded
by a light meson cloud. Such fields are usually classified by the 
value\footnote{ Actually if we want to picture the linked mesons as 
literally composed of a meson--meson pair, we should assign relative 
orbital angular momentum $\ell-1$ to these bosonic constituents and
allow for both light pseudoscalars and vectors.}, $\ell$ of the
relative orbital angular momentum of a $\bar{q}Q$ pair which describes
it in the CQM.  
We will not attempt to explain the binding of these two  
mesons but shall simply incorporate the ``experimental'' higher spin 
meson fields into our chiral Lagrangian. Different $\ell$ excitations
will correspond to the use of different meson field multiplets.
{}From now on we will restrict our attention to $N_{\rm C}=3$.

Taking the new degree of freedom $\mbox{\boldmath$\ell$}$ into account
requires us to modify the previous formulas describing the heavy
baryon.
Now the parity formula (\ref{baryon parity}) is modified to 
\begin{equation}
P_B = \left( -1 \right)^{\ell+r} \ ,
\label{new baryon parity}
\end{equation}
which is seen to be compatible with the CQM relation 
(\ref{baryon parity:CQM}).
Now Eq.~(\ref{baryon spin}) holds but with the light grand spin 
$\mbox{\boldmath$g$}$ modified to,
\begin{equation}
\mbox{\boldmath$g$} = 
\mbox{\boldmath$r$} + \mbox{\boldmath$K'$} \ .
\label{def: g}
\end{equation}
Note that $\mbox{\boldmath$K$}$ in Eq.~(\ref{def: k}) has been
incorporated in
\begin{equation}
\mbox{\boldmath$K'$} = \mbox{\boldmath$I$}_{\rm light} +
\mbox{\boldmath$S$}_{\rm light} + \mbox{\boldmath$\ell$}
\ .
\label{def: k prime}
\end{equation}
The new correspondence between the bound--state picture variables and
those of the CQM is:
\begin{eqnarray}
\mbox{\boldmath$S$}_{\rm heavy}
&\leftrightarrow&
\mbox{\boldmath$S$}_H
\ ,
\nonumber\\
\mbox{\boldmath$r$}
&\leftrightarrow&
\mbox{\boldmath$L$}_E
\ ,
\nonumber\\
\mbox{\boldmath$\ell$}
&\leftrightarrow&
\mbox{\boldmath$L$}_I
\ ,
\nonumber\\
\mbox{\boldmath$I$}_{\rm light} +
\mbox{\boldmath$S$}_{\rm light} +
\mbox{\boldmath$J$}^{\rm sol}
&\leftrightarrow&
\mbox{\boldmath$S$}
\ .
\label{correspondence:2}
\end{eqnarray}
Previously $\mbox{\boldmath$I$}_{\rm light} + 
\mbox{\boldmath$S$}_{\rm light} = \mbox{\boldmath$K$}$ had zero
quantum numbers on the bound states; now the picture is a little more
complicated.
We will see that the dynamics may lead to new bound states which are in
correspondence with the CQM.
Equation~(\ref{correspondence:2}) should be interpreted in the sense
of this correspondence.

It is easiest to see that the lowest new states generated agree with
the CQM for $\ell=\mbox{even}$, which corresponds to negative parity
heavy mesons. In this case $k=0$ or equivalently $k'=\ell$ may be 
favored dynamically. Then the last line in Eq.~(\ref{correspondence:2}) 
indicates that $J^{\rm sol}$, which can take on the 
values $0$ and $1$, corresponds to the light diquark spin $S$ in the CQM.
This leads to the CQM states of type a) and b) in Eq.~(\ref{condition:CQM}).
This is just a generalization of the discussion for the ground state
given in section~\ref{sec:preliminaries}. Now let us discuss how 
the states corresponding to c) and d) can be constructed in the bound 
state scenario. Apparently we require $\ell=\mbox{odd}$, {\it i.e.} 
positive parity heavy mesons. For $I=0$ we also have $J^{\rm sol}=0$. 
Hence the last line in Eq.~(\ref{correspondence:2}) requires $k=1$ 
for $S=1$. To generate states of type d) also $k=1$ would be needed in 
order to accommodate $I=J^{\rm sol}=1$ and $S=0$. Actually for the 
case $k=1$ and $J^{\rm sol}=1$ states with $S=0,1,2$ would 
be possible. The states with $S=1,2$ should be ruled out by the 
dynamics of the model.

One may perhaps wonder whether we are pushing the bound state picture
too far; since things seen to be getting more complicated why not just
use the constituent quark model?
Apart from the intrinsic interest of the soliton approach there are
two more or less practical reasons for pursuing the approach.
The first is that the parameters of the underlying chiral Lagrangian
are, unlike parameters such as the constituent quark masses and
inter--quark potentials of the CQM, physical ones and in principle
subject to direct experimental test.
The second reason is that the bound state approach actually models the
expected behavior of a confining theory; namely, when sufficient
energy is applied to ``stretch'' the heavy baryon it does not come
apart into a heavy quark and two light quarks but rather into a
nucleon and a heavy meson.
The light quark--antiquark pair which one usually imagines popping out
of the vacuum when the color singlet state has been
suitably stretched, was there all the time, waiting to play a role, in
the bound state picture.
The model may therefore be useful in treating reactions of this sort.

\section{A model for the missing first excited states}
\label{sec:4}

Before going on to the general orbital excited states it may be
helpful to 
see how the dynamics could work out for explaining the missing seven 
$\Lambda_{Q}$ 
and $\Sigma_{Q}$ type, negative parity, excited states. In the {\it new} 
bound state picture these correspond to the choices\footnote{%
Actually, $\ell$ was 
introduced for convenience in making a comparison with the
constituent quark  
model. It is really hidden in the heavy mesons which, strictly
speaking, are  
specified by the light cloud angular momentum 
$\mbox{\boldmath$J$}_{\rm light}$ 
and parity. We can perform the calculation without mentioning $\ell$.
} 
$\ell=1$, $r=0$. As discussed, we are considering that the orbital
angular momentum $\ell$ is ``locked--up''
in suitable excited heavy mesons. As in Eq.~(\ref{def:wave function}),
$r$ appears as a parameter in the new heavy meson wave--function. The 
treatment of the excited heavy mesons in the effective theory context,
has been given already by Falk and Luke~\cite{Falk-Luke}. For a 
review see \cite{Ca97}. The case (for orbital angular 
momentum=1) where the light cloud spin of the heavy meson is $1/2$ is 
described by the heavy multiplet 
\begin{equation}
{\cal{H}} = \frac{1 - i \gamma_\mu V_\mu}{2}
\left(S + i \gamma_{5}\gamma_\nu A_\nu  \right) \ ,
\label{def:curH}
\end{equation}
where $S$ is the fluctuation field for a scalar ($J^P=0^+$) 
particle and $A_{\mu}$, satisfying $V_{\mu}A_{\mu}=0$, similarly corresponds 
to an axial ($J^P=1^+$) particle. The case where the light cloud 
spin is $3/2$ is described by 
\begin{equation}
{\cal{H_{\mu}}}=
\frac{1 - i \gamma_\alpha V_\alpha}{2}\left(-T_{\mu\nu}\gamma_{\nu} +
i \sqrt{\frac{3}{2}}B_{\nu}\gamma_5\left[\delta_{\mu\nu} -
\frac{1}{3}\gamma_{\nu}  
\left(\gamma_{\mu}+iV_{\mu}\right)\right]\right)
\label{def:curHm}
\end{equation}
satisfying the {\it Rarita-Schwinger} constraints 
${\cal H}_{\mu}\gamma_{\mu}={\cal H}_{\mu} V_{\mu}=0$. The field 
$T_{\mu\nu}=T_{\nu\mu}$ (with $V_{\mu}T_{\mu\nu}=T_{\mu\mu}=0$) is a
spin $2$  
tensor ($J^P=2^+$) and $B_{\mu}$ (with $V_{\mu}B_{\mu}=0$) is 
another axial ($J^P=1^+$). Currently, experimental
candidates exist  
for the tensor and an axial.

In order to prevent the calculation from becoming too complicated we
will, for the purpose of the present paper, adopt the approximation of
leaving out the light vector mesons.
This is a common approximation used by workers in the field but it
should be kept in mind that the effect of the light vectors is
expected to be substantial.

The kinetic terms of the effective chiral Lagrangian (analogous
to the  first term of Eq.~(\ref{Lag for H})) are:
\begin{equation}
{\cal L}_{\rm kin}
= -i {\cal M}V_{\mu}\mbox{Tr}\, 
\left[ {\cal H} D_\mu \bar{{\cal H}} \right]
+ i {\cal M} \, V_{\mu}\mbox{Tr}\, 
\left[ {\cal H}_{\mu} D_\mu \bar{\cal H}_{\mu}
\right] \ ,
\label{Lag for curlH}
\end{equation} 
where ${\cal M}$ is a characteristic heavy mass scale for the excited
mesons.  
For simplicity\footnote{%
A more general approach is to replace ${\cal M}$ on the 
 right hand side of Eq.~(\ref{Lag for curlH}) by the same ${M}$ used
in Eq.~(\ref{Lag for H}) and to add the splitting terms 
$-2M(M_S-M)\mbox{Tr}\,\left[ {\cal H}\bar{\cal H}\right] + 
2M(M_T-M)\mbox{Tr}\,\left[{\cal H}_{\mu}\bar{\cal H}_{\mu}\right]$.} 
we are neglecting mass differences between the $\ell=1$ heavy mesons.
The interaction terms involving only the ${\cal H}$ and 
${\cal H}_{\mu}$ fields,  to lowest order in derivatives, are 
\begin{eqnarray}
{\cal L}_{\rm int}/{\cal M}&=& i d_{\rm S} \mbox{Tr}\,
\left[{\cal H}\gamma_{\mu}\gamma_5 p_{\mu}\bar{\cal H}\right] 
- i d_{\rm T} \mbox{Tr}\, \left[{\cal H}_{\mu}\gamma_{\alpha}
\gamma_5 p_{\alpha}\bar{\cal H}_{\mu}\right] \nonumber \\
 &+&\left[
i f_{\rm ST}\mbox{Tr}\,
\left[{\cal H}\gamma_{5} p_{\mu}\bar{\cal H}_{\mu}\right] + 
\mbox{h.c.}
\right] \ . 
\label{Int for curlH}
\end{eqnarray}
These generalize the second term in Eq.~(\ref{Lag for H}) and 
$d_{\rm S}$, $d_{\rm T}$  
 and $f_{\rm ST}$ (which may be complex) are the heavy meson--pion
coupling  
constants. Similar terms which involve $\ell\neq 1$ multiplets are not
needed  
for our present purpose but will be discussed in the next section.

As in  section \ref{sec:preliminaries},
the wave--functions for the excited heavy 
mesons bound to the background Skyrmion are conveniently presented in
the rest frame $\mbox{\boldmath$V$}=0$. 
The analogs of Eq.~(\ref{H: matrix}) become  
\begin{equation}
\bar{{\cal H}}_{\rm c} \rightarrow
\left( \begin{array}{cc}
\bar{f}_{lh}^a & 0 \\ 
0 & 0 
\end{array}\right) 
\ , \quad
\left(\bar{{\cal H}}_{i}\right)_{\rm c} \rightarrow
\left( \begin{array}{cc}
0 & 0 \\ \bar{f}_{i,lh}^a & 0 
\end{array}\right)
\ ,
\label{curlH: matrix}
\end{equation}
and $\left(\bar{{\cal H}}_0\right)_{\rm c}\rightarrow 0$. Now the 
wave--functions in Eq.~(\ref{curlH: matrix}) are expanded as:
\begin{eqnarray}
\bar{f}_{lh}^a &=& 
\frac{u\left( \vert\mbox{\boldmath$x$}\vert\right)}{\sqrt{{\cal M}}} 
\left(\hat{\mbox{\boldmath$x$}}\cdot
\mbox{\boldmath$\tau$}_{ad}\right) 
\Phi_{ld}\left(k^{\prime},k^{\prime}_3;r\right) 
\chi_{h} \ , \nonumber \\
\bar{f}_{i,lh}^a &=& 
\frac{u\left( \vert\mbox{\boldmath$x$}\vert\right)}{\sqrt{{\cal M}}} 
\left(\hat{\mbox{\boldmath$x$}}\cdot
\mbox{\boldmath$\tau$}_{ad}\right)\Phi_{i,ld}
\left(k^{\prime},k^{\prime}_3;r\right)\chi_{h} \ ,
\label{f: classical}
\end{eqnarray}
where $u$ stands for a sharply peaked radial wave--function which 
may differ for the two cases. Other notations are as in 
Eq.~(\ref{h: classical}). Note that the constraint 
$\gamma_{\mu}\bar{\cal H}_{\mu}=0$ implies that 
\begin{equation}
\left(\sigma_i\right)_{ll'}\Phi_{i,l'd}=0  \ .
\label{rarita}
\end{equation}
It is interesting to see explicitly how the extra angular 
momentum $\ell=1$ is ``locked--up'' in the heavy meson
wave--functions.  
For the ${\cal H}$ wave--function, the fact that 
$\mbox{\boldmath $J$}_{\rm light}=\mbox{\boldmath$\ell$}+
\mbox{\boldmath$S$}_{\rm light}$ 
takes the value $1/2$ leads, using Eq.~(\ref{def: k prime}), 
to the possible 
values $k^{\prime}=0$ or $1$. The corresponding wave--functions are 
\begin{equation}
\Phi_{ld}\left(k^{\prime}=k^{\prime}_3=0\right)=
\frac{\epsilon_{ld}}{\sqrt{8\pi}}\ , 
\quad
\Phi_{ld}\left(k^{\prime}=k_3=1\right)=
\frac{\delta_{l1}\delta_{d1}}{\sqrt{4\pi}}\ , 
\label{curlH wave--function}
\end{equation}
where, for the present case, we are taking $r=0$. 
For the $\bar{\cal H}_i$ 
wave--function it is important to satisfy 
$j_l=\vert\mbox{\boldmath$J$}_{\rm light}\vert$ = $3/2$ condition 
(\ref{rarita}). This may be accomplished by combining with suitable 
Clebsch-Gordan coefficients an $\ell=1$ wave--function with the 
$S_{\rm light}=1/2$ 
spinor to give 
\begin{eqnarray}
\Phi_{i,ld}\left(k^{\prime}=k^{\prime}_3=2\right)&=&w^{(+1)}_i \delta_{l1} 
\delta_{d1} \ , \nonumber \\
\Phi_{i,ld}\left(k^{\prime}=k^{\prime}_3=1\right)&=& 
\frac{\sqrt{3}}{2}w^{(+1)}_i \delta_{l1} \delta_{d2} - \frac{1}{2\sqrt{3}} 
w^{(+1)}_{i} \delta_{l2}\delta_{d1} - \frac{1}{\sqrt{6}} w^{(0)}_{i}
\delta_{l1}\delta_{d1} 
\ , 
\label{curlHmu wave--function}
\end{eqnarray}
where 
$w^{(\pm 1)}_{j}=\frac{\mp 1}{\sqrt{8 \pi}} 
\left(\delta_{j1}\pm i\delta_{j2}\right)$ and 
$w^{(0)}_i=\frac{\delta_{i3}}{\sqrt{4\pi}}$ is a spherical
decomposition.

The main question is: 
Which of the channels contain bound states?
Note that, for the reduced space in which
$\hat{\mbox{\boldmath$x$}}\cdot \mbox{\boldmath$\tau$}$ 
has been removed 
as in Eq.~(\ref{f: classical}), 
$k^{\prime}$ is a good quantum number. 
Furthermore, because the wave--function 
$u\left(\vert\mbox{\boldmath$x$}\vert\right)$ is sharply peaked,
the relevant matrix elements are actually 
independent of the orbital angular momentum $r$.
The classical potential for each $k^{\prime}$ channel may be
calculated  by setting $r=0$ and substituting the appropriate reduced
wave--functions from Eqs.~(\ref{curlH wave--function}) and 
(\ref{curlHmu wave--function}) into the  
interaction Lagrangian (\ref{Int for curlH}). 
(see Appendix~\ref{app:a} for more details.)
The $k^{\prime}=0$ channel 
gets a contribution only from the $d_{\rm S}$ term in 
Eq.~(\ref{Int for curlH}) 
while the $k^{\prime}=2$ channel receives a contribution only from 
the $d_{\rm T}$ term. On the other hand, all three terms contribute to
the  
$k^{\prime}=1$ channel. The resulting potentials are:
\begin{eqnarray}
V\left(k^{\prime}=0\right)&=& - \frac{3}{2}d_{\rm S} F^{\prime}(0) 
\label{kprime=0}\ , \\
V\left(k^{\prime}=2\right)&=& - \frac{1}{2}d_{\rm T} F^{\prime}(0) 
\label{kprime=2}\ , 
\end{eqnarray}
\begin{equation}
V\left(k^{\prime}=1\right)=
\left( \begin{array}{cc}
\langle{\cal H }|V|{\cal H}\rangle & 
 \langle{\cal H }|V|{\cal H}_{\mu}\rangle \\ 
\langle{\cal H}_{\mu} |V| {\cal H}\rangle  &  
 \langle{\cal H}_{\mu} |V| {\cal H}_{\mu}\rangle
\end{array}\right)
= 
\left( \begin{array}{cc}
\frac{1}{2}d_{\rm S} & - i \sqrt{\frac{2}{3}}f_{\rm ST} \\ 
 i \sqrt{\frac{2}{3}}f_{\rm ST}^*  & \frac{5}{6} d_{\rm T} 
\end{array}\right) F'(0)
\ .
\label{kprime=1}
\end{equation}
The classical criterion for a channel to contain a bound state is 
that its potential be negative. Since $F^{\prime}(0)>0$ we require for 
bound states in the $k^{\prime}=0$ and $k^{\prime}=2$ channels 
\begin{equation}
d_{\rm S} > 0 \ , \quad d_{\rm T}>0 \ ,
\label{dst constraints}
\end{equation}
respectively\footnote{
In a more general picture where $\ell=3$ excited heavy 
mesons are included, the $k^{\prime}=2$ channel will also be described
by a potential matrix. Then the criterion for $d_{\rm T}$ is modified.
(See next section.)
}. 
For bound 
states in the $k^{\prime}=1$ channel we must examine the signs of the 
eigenvalues of Eq.~(\ref{kprime=1}).  Assuming that 
Eq.~(\ref{dst constraints}) holds (as will be seen to be desirable) it
is easy to see that there is, at most, {\it one} 
$k^{\prime}=1$ bound state. The condition  
for this bound state to exist is 
\begin{equation}
\left\vert{f_{\rm ST}}\right\vert^2 > \frac{5}{8} \, d_{S}\, d_{T} \ .
\label{fst constraints}
\end{equation}
The (primed) states which diagonalize Eq.~(\ref{kprime=1}) are simply 
related to the original ones by 
\begin{equation}
\left( \begin{array}{c}
\Phi\\ 
 \Phi_{i}
\end{array}\right)=
\left( \begin{array}{cc}
\cos\theta&\sin\theta\\ 
-i p^* \sin\theta& i p \cos\theta
\end{array}\right)
\left( \begin{array}{c}
\Phi^{\prime}\\ 
 \Phi_{i}^{\prime}
\end{array}\right) \ ,
\label{rotation}
\end{equation}
\begin{equation}
\tan2\theta=
\frac{4 \sqrt{6} \left\vert f_{\rm ST}\right\vert}{
5 d_{\rm T} - 3 d_{\rm S}} \ ,
\label{def: theta}
\end{equation}
where $p$ is the phase of $f_{\rm ST}$. $\Phi$ and 
$\Phi_{i}$ are shorthand 
notations\footnote{Strictly speaking, to put $\Phi_{ld}$ on a 
parallel footing to $\Phi_{i,ld}$ we should replace 
 $\Phi_{ld} \rightarrow 
 \sqrt{\frac{3}{8}}\left(P^{3/2}\right)_{ik;ll^{\prime}} 
 \left(\tau_{k}\right)_{dd^{\prime}}\Phi_{l^{\prime}d^{\prime}}$ with
the spin $3/2$ projection operator, 
 $\left(P^{3/2}\right)_{ik;ll^{\prime}}
 = \frac{2}{3}\left(\delta_{ik} \delta_{ll^{\prime}} - 
 \frac{i}{2}\epsilon_{jik}
 \left(\sigma_{j}\right)_{ll^{\prime}}\right)$
 (see Appendix~\ref{app:a}).
}
for the appropriate wave--functions.
Clearly, the results for which states are bound depend on the numerical 
values and signs of the coupling constants. At the moment there is 
no purely experimental information on these quantities. However, it is
 very interesting to observe that if Eqs.~(\ref{dst constraints}) and
(\ref{fst constraints}) hold, 
then the missing first excited $\Lambda_{Q}$ 
states are bound. To see this note that the heavy baryon spin is given 
by Eq.~(\ref{baryon spin}) with $\mbox{\boldmath$g$}$ defined 
in Eqs.~(\ref{def: g}) and 
(\ref{def: k prime}). For the $\Lambda_{Q}$--type states, noting 
that $I=J^{\rm sol}=0$ in the Skyrme approach gives the baryon spin as 
\begin{equation}
\mbox{\boldmath$J$}=\mbox{\boldmath$g$}+
\mbox{\boldmath$S$}_{\rm heavy}
\quad
\left(\Lambda_{Q}~{\rm states}\right) \ .
\label{J baryon}
\end{equation}
The $r=0$ choice enables us to set $g=k^{\prime}$. With just the 
three attractive channels $k^{\prime}=0$, 
$k^{\prime}=1$ and $k^{\prime}=2$ we 
thus end up with the missing first three excited $\Lambda_{Q}$ heavy 
multiplets $\Lambda_{Q}\left(\frac{1}{2}^{-}\right)$ ,
 $\left\{\Lambda_{Q}\left(\frac{1}{2}^{-}\right), 
\Lambda_{Q}\left(\frac{3}{2}^{-}\right)\right\}$ and 
 $\left\{\Lambda_{Q}\left(\frac{3}{2}^{-}\right), 
\Lambda_{Q}\left(\frac{5}{2}^{-}\right)\right\}$. 
It should be stressed that this counting involves dynamics rather than
pure  
kinematics. For example, it may be seen from
Eqs.~(\ref{kprime=0})--(\ref{kprime=1})  
that it is dynamically impossible to have four bound heavy multiplets 
($k^{\prime}=0,~ k^{\prime}=2$ and two $k^{\prime}=1$ channels).  
The missing first excited $\Sigma_{Q}$--type states comprise the
single  heavy multiplet 
 $\left\{\Sigma_{Q}\left(\frac{1}{2}^{-}\right), 
\Sigma_{Q}\left(\frac{3}{2}^{-}\right)\right\}$. At 
the classical level there are apparently more bound multiplets present. 
However, we will now see that the introduction of collective
coordinates,  
as is anyway required in the Skyrme model~\cite{Sk61}
to generate states with good 
isospin quantum number, will split the heavy multiplets from each other. 
Thus, deciding which states are bound actually requires 
a more detailed analysis.

We need to extend Eq..~(\ref{def: collective}) 
in order to allow the $\ell=1$ heavy meson 
fields to depend on the collective rotation variable $A(t)$:
\begin{equation}
\bar{{\cal H}}(\mbox{\boldmath$x$},t)=
A(t) \bar{{\cal H}}_{\rm c} (\mbox{\boldmath$x$}) 
\ ,
\quad
\bar{{\cal H}}_{i}(\mbox{\boldmath$x$},t)=A(t)
\bar{{\cal H}}_{i{\rm c}} (\mbox{\boldmath$x$}) 
\ ,
\label{def: curl collective}
\end{equation}
where $\bar{{\cal H}}_{\rm c}$ and $\bar{{\cal H}}_{i{\rm c}}$ 
are given in Eq.~(\ref{curlH: matrix}). 
Note, again, that the matrix $A(t)$ acts on 
the isospin indices. We also have $\bar{{\cal H}}_{0{\rm c}}=0$ 
due to the 
rest frame constraint $V_{\mu}\bar{{\cal H}}_{\mu {\rm c}}=0$. 
Now substituting Eq.~(\ref{def: curl collective}) as well as the first
of Eq.~(\ref{def: collective}) into the heavy field 
Lagrangian\footnote{%
Note that Eq.~(\ref{Lag for curlH}) contributes but 
Eq.~(\ref{Int for curlH}) does not contribute.} 
yields~\cite{Callan-Klebanov} the collective Lagrangian\footnote{%
In Eq.~(\ref{def: Lcoll}) $k^{\prime}$ is defined to 
operate on the heavy particle wave--functions rather than on their
conjugates.  
This is required when the heavy meson is coupled to the Skyrme
background field since $\Lambda_{Q}$ is made as 
$\left(qqq\right)\left(\bar{q}Q\right)$ 
rather than $\left(qqq\right)\left(\bar{Q}q\right)$. For convenience 
in Eqs.~(\ref{h: classical}) and (\ref{f: classical}) we have 
considered the conjugate wave--functions 
(since they are usual in the light sector). This has been compensated 
by the minus sign in the second term of Eq.~(\ref{def: Lcoll}).}
\begin{equation} 
L_{\rm coll}=\frac{1}{2}\,\alpha^2 \mbox{\boldmath$\Omega$}^2 - 
\chi\left(k^{\prime}\right) 
\mbox{\boldmath$K^{\prime}$}\cdot\mbox{\boldmath$\Omega$} \ ,
\label{def: Lcoll}
\end{equation}
where $\mbox{\boldmath$\Omega$}$ is defined in Eq.~(\ref{def: omega})
and  $\alpha^2$ is the Skyrme model moment of inertia. In the vector 
meson model the induced fields ($\rho^a_0$ and $\omega_i$) are 
determined from a variational approach to $\alpha^2$. The 
quantities $\chi\left(k^{\prime}\right)$ are given by 
(see Appendix~\ref{app:b}).
\begin{equation}
\chi\left(k^{\prime}\right)=
\left\{\begin{array}{ll}
0 & k^{\prime}=0 \\
\frac{1}{4}\left(3\,\cos^2\theta - 1\right) 
& k^{\prime}=1 \\
\frac{1}{4} & k^{\prime}=2 
\end{array}\right. \ ,
\label{def: chi}
\end{equation}
where the angle $\theta$ is defined in Eq.~(\ref{def: theta}). 
(Note that if light vector mesons are included the expressions
for $\chi$ would be more involved as the induced fields will also
contribute.)
In writing Eq.~(\ref{def: chi}) it was assumed that the first state in
Eq.~(\ref{rotation}) 
(i.e. $\Phi^{\prime}$ rather than $\Phi^{\prime}_i$) is the bound one;
the collective Lagrangian is constructed as an expansion around the
bound 
state solutions. We next determine from Eq.~(\ref{Jsol}), the
canonical (angular)  
momentum $\mbox{\boldmath$J$}^{\rm sol}$ as 
$\alpha^2\,\mbox{\boldmath$\Omega$} - \chi\left(k^{\prime}\right)
 \mbox{\boldmath$K$}^{\prime}$. The usual Legendre transform then
leads to the collective Hamiltonian
\begin{equation}
H_{\rm coll}=\frac{1}{2\alpha^2} \left(\mbox{\boldmath$J$}^{\rm sol}+
\chi\left(k^{\prime}\right)\,\mbox{\boldmath$K^{\prime}$}\right)^2 \ .
\label{Hcoll}
\end{equation}
Again we remark that 
$J^{\rm sol}=I$. 
It is useful to define the light part of the total heavy baryon spin
as  
\begin{equation}
\mbox{\boldmath$j$}= \mbox{\boldmath$r$} + 
\mbox{\boldmath$K$}^{\prime}+\mbox{\boldmath$J$}^{\rm sol} \ ,
\label{baryon jlight}
\end{equation}
and rewrite Eq.~(\ref{Hcoll}) as 
\begin{equation}
H_{\rm coll}=\frac{1}{2\alpha^2}\left[
\left(1 - \chi\left(k^{\prime}\right)\right)\mbox{\boldmath$I$}^2 + 
 \chi\left(k^{\prime}\right)\left(\mbox{\boldmath$j$} - 
\mbox{\boldmath$r$}\right)^2
+ \chi\left(k^{\prime}\right)\left(\chi\left(k^{\prime}\right) 
- 1 \right)
\mbox{\boldmath$K^{\prime}$}^2 \right] \ . 
\label{Hcoll 2}
\end{equation}
The mass splittings within each given 
$k^{\prime}$ multiplet due to $H_{\rm coll}$ 
are displayed in Table~\ref{tab:1}.
\begin{table}[htbp]
\begin{center}
\small
\begin{tabular}{c|c|c|c|c|c}
 $I$ & $k^{\prime}$ 
 & $\left\vert
      \mbox{\boldmath$K$}^\prime+\mbox{\boldmath$J$}^{\rm sol}
   \right\vert$
 & $V$ & $\alpha^2 H_{\rm coll}$ 
&{Candidates for $r=0$} \\
 $=J^{\rm sol}$ &~ &~ &~ &~ &  {missing states}~~~~~ \\
\hline
 & 0 & 0 & $-\frac{3}{2} d_{\rm S}\, F^{\prime}(0)$ & 0 
& $\Lambda_{Q}\left(\frac{1}{2}^{-}\right)$\\
~{\Large 0}~ & 1 & 1  & $\lambda$ & $\chi^2$ 
&  $\left\{\Lambda_{Q}\left(\frac{1}{2}^{-}\right), 
\Lambda_{Q}\left(\frac{3}{2}^{-}\right)\right\}$\\
 & 2 & 2  & $-\frac{1}{2}d_{\rm T}\, F^{\prime}(0) $ & $\frac{3}{16}$ 
&  $\left\{\Lambda_{Q}\left(\frac{3}{2}^{-}\right), 
\Lambda_{Q}\left(\frac{5}{2}^{-}\right)\right\}$\\
\hline
 & 0  &  1  & $-\frac{3}{2}d_{\rm S}\, F^{\prime}(0) $ & 1  
&  $\left\{\Sigma_{Q}\left(\frac{1}{2}^{-}\right), 
\Sigma^{\prime}_{Q}\left(\frac{3}{2}^{-}\right)\right\}_{1}$\\
 & 1  &  0  & $\lambda $ & $\left(\chi - 1\right)^2$ & \\
 & 1  &  1  & $''$& $\left(\chi - 1\right)^2 + \chi$
&  $\left\{\Sigma_{Q}\left(\frac{1}{2}^{-}\right), 
\Sigma_{Q}\left(\frac{3}{2}^{-}\right)\right\}_{2}$\\
~{\Large 1}~ & 1  &  2  & $''$& $\left(\chi - 1\right)^2 + 3\,\chi$ & \\
 & 2  &  1  & $-\frac{1}{2}d_{\rm T}\,F^{\prime}(0)$ & $\frac{7}{16}$
&  $\left\{\Sigma_{Q}\left(\frac{1}{2}^{-}\right), 
\Sigma_{Q}\left(\frac{3}{2}^{-}\right)\right\}_{3}$\\
 & 2  &  2  & $''$ & $\frac{15}{16}$ &  \\
 & 2  &  3  & $''$ & $\frac{27}{16}$ &  \\
\end{tabular}
\end{center}
\caption[]{
Contributions to energies of new predicted $\ell=1$ states.
Here, 
$\lambda=\frac{1}{4}F^{\prime}(0) 
\left[\left(d_{\rm S} + \frac{5}{3} d_{\rm T} \right) - 
\sqrt{\left(d_{\rm S} - \frac{5}{3} d_{\rm T} \right)^2 +
\frac{32}{3} |f_{\rm ST}|^2} \right]$ is the presumed negative binding 
potential 
in the $k^{\prime}=1$ channel. Furthermore $\chi=\chi(1)$ in
Eq.~(\ref{def: chi});  
it satisfies $-\frac{1}{4}\leq \chi \leq \frac{1}{2}$.
} 
\label{tab:1}
\end{table}
\noindent
This table also shows the splitting of the $k^{\prime}$ multiplets from 
each other due to the classical potential in 
Eqs.~(\ref{kprime=0})--(\ref{kprime=1}). Note that the slope of the 
Skyrme profile function $F^{\prime}(0)$ is of order $1$ GeV. The 
coupling constants $d_{\rm S}, d_{\rm T}, f_{\rm ST}$, 
based on $d\simeq 0.5$ for the 
ground state heavy meson, are expected to be of the order unity.
Hence the binding potentials $V$ are expected to 
be of the rough order of $500$ MeV. The inverse moment of inertia 
 $1/\alpha^2$ is of the order of $200$ MeV which 
(together with $-\frac{1}{4}\leq \chi \leq \frac{1}{2}$) sets the scale 
for the ``$1/N_{\rm C}$'' corrections due to $H_{\rm coll}$. 
As mentioned before, 
if the coupling constants satisfy the inequalities 
(\ref{dst constraints}) and (\ref{fst constraints}), all the 
$\Lambda_{Q}$ multiplets shown will be bound. At first  
glance we might expect all the $\Sigma_{Q}$ states listed 
also to be bound. However the $H_{\rm coll}$ corrections increase as
$I$ 
increases, which is a possible indication that many of the 
$\Sigma_{Q}$'s might be only weakly bound. In a more complete 
model they may become unbound. Hence it is interesting to ask which 
of the three displayed candidates for the single missing 
$\Sigma_{Q}$ multiplet is mostly tightly bound in the present 
model. Neglecting the effect of $V$ we can see that $H_{\rm coll}$ 
raises the  energy of candidate 3 less than those of candidates 1  
and 2. Furthermore, for the large range of $\chi$, 
$-\frac{1}{4}\leq \chi \leq 1 - \frac{\sqrt{7}}{4}$, candidate 3 suffers 
the least unbinding due to $H_{\rm coll}$ of any of the $I=1$ heavy
baryons  listed. 
The $\Lambda_{Q}$ states suffer still less unbinding
due  
to $H_{\rm coll}$.

\section{Higher Orbital Excitations}
\label{sec:5}

We have already explicitly seen that the ``missing'' first orbitally
excited heavy baryon states in the bound state picture might be
generated if the model is extended to also include binding the first 
orbitally excited heavy mesons in the background field of a Skyrme 
soliton. From the correspondence (\ref{correspondence:2}) and 
associated discussion we expect that any of the higher excited heavy 
baryons of the CQM might be similarly generated by binding the 
appropriately excited heavy mesons. In this section we will show 
in detail how this result can be achieved in the general case.
An extra complication, which was neglected for simplicity in the last 
section, is the possibility of baryon states constructed by binding
heavy mesons of different $\ell$, mixing with each other.
For example $\{r=1\,,\,\ell=0\}$ 
type states can mix with $\{r=1\,,\,\ell=2\}$ 
type states, other quantum numbers being the same.
Since $\mbox{\boldmath$r$}+\mbox{\boldmath$\ell$}$ must add to $1$, 
this channel could not mix with $\{r=1\,,\,\ell=4\}$.
An identical type of mixing -- between $\{L_E=1\,,\,L_I=0\}$ and
$\{L_E=1\,,\,L_I=2\}$ -- may also exist in the CQM. The present model, 
however, provides a simple way to study this kind of mixing as a 
perturbation.

\begin{table}[htbp]
\begin{center}
\begin{tabular}{cccc}
field & $\ell$ & $j_l$ & $J^P$ \\
\hline
$H$ & 0 & $1/2$ & $0^-$, $1^-$ \\
& & & \\
${\cal H}$ & 1 & $1/2$ & $0^+$, $1^+$ \\
${\cal H}_\mu$ & 1 & $3/2$ & $1^+$, $2^+$ \\
& & & \\
$H_\mu$ & 2 & $3/2$ & $1^-$, $2^-$ \\
$H_{\mu\nu}$ & 2 & $5/2$ & $2^-$, $3^-$ \\
& & & \\
\vdots & & & \\
& & & \\
$H_{\mu_1\cdots\mu_{\ell-1}}$ & $\ell=\mbox{even}$
 & $\ell-1/2$ & $(\ell-1)^-$, $\ell^-$ \\
$H_{\mu_1\cdots\mu_\ell}$ & $\ell=\mbox{even}$
 & $\ell+1/2$ & $\ell^-$, $(\ell+1)^-$ \\
& & & \\
${\cal H}_{\mu_1\cdots\mu_{\ell-1}}$ & $\ell=\mbox{odd}$
 & $\ell-1/2$ & $(\ell-1)^+$, $\ell^+$ \\
${\cal H}_{\mu_1\cdots\mu_\ell}$ & $\ell=\mbox{odd}$
 & $\ell+1/2$ & $\ell^+$, $(\ell+1)^+$ \\
& & & \\
\vdots & & & \\
\end{tabular}
\end{center}
\caption[]{%
Notation for the heavy meson multiplets.
$j_l$ is the angular momentum of the ``light cloud'' surrounding the
heavy quark while $J^P$ is the spin parity of each heavy meson in the
multiplet.
}
\label{table:3}
\end{table}
To start the analysis it may be helpful to refer to
Table~\ref{table:3}, which shows our notations for the excited heavy
meson multiplet  ``fluctuation'' fields. 
The straight $H$'s contain negative parity mesons and 
the curly ${\cal H}$'s contain positive parity mesons. 
Further details are given 
in Ref.~\cite{Falk-Luke}. Note that each field is symmetric in 
all Lorentz indices and obeys the constraints
\begin{equation}
V_{\mu_1} H_{\mu_1\cdots\mu_n} = 
H_{\mu_1\cdots\mu_n} \gamma_{\mu_1} = 0 \ ,
\end{equation}
as well as for ${\cal H}_{\mu_1\cdots\mu_n}$.
The general chiral invariant interaction with the lowest number of 
derivatives is
\begin{equation}
{\cal L}_{\rm d} + {\cal L}_{\rm f} + {\cal L}_{\rm g} \ ,
\end{equation}
where
\begin{eqnarray}
{\cal L}_{\rm d} &=& i M \sum_{n=0} d_{{\rm P}n} \left(-1\right)^n
\mbox{\rm Tr}\, 
\left[
  H_{\mu_1\cdots\mu_n} p_\mu \gamma_\mu \gamma_5 
  \bar{H}_{\mu_1\cdots\mu_n}
\right]
\nonumber\\
&& {}
+ i M \sum_{n=0} d_{{\rm S}n} \left(-1\right)^n
\mbox{\rm Tr}\, 
\left[
  {\cal H}_{\mu_1\cdots\mu_n} p_\mu \gamma_\mu \gamma_5 
  \bar{\cal H}_{\mu_1\cdots\mu_n}
\right]
\ ,
\nonumber\\
{\cal L}_{\rm f} &=& i M \sum_{n=0} f_{{\rm P}n} \left(-1\right)^n
\mbox{\rm Tr}\, 
\left[
  H_{\mu_1\cdots\mu_n} p_\mu \gamma_5 \bar{H}_{\mu_1\cdots\mu_n\mu}
\right]
+ \mbox{h.c.}
\nonumber\\
&& {}
+ i M \sum_{n=0} f_{{\rm S}n} \left(-1\right)^n \mbox{\rm Tr}\, 
\left[
  {\cal H}_{\mu_1\cdots\mu_n} p_\mu \gamma_5 
  \bar{\cal H}_{\mu_1\cdots\mu_n\mu}
\right]
+ \mbox{h.c.}
\ .
\label{general d f term}
\end{eqnarray}
The final piece,
\begin{equation}
{\cal L}_{\rm g} = i M \sum_{n=0} g_{n} \left(-1\right)^n
\mbox{\rm Tr}\, 
\left[
  {\cal H}_{\mu_1\cdots\mu_n} p_\mu \gamma_\mu \gamma_5 
  \bar{H}_{\mu_1\cdots\mu_n}
\right]
+ \mbox{h.c.}
\label{general g term}
\end{equation}
exists in general, but does not contribute for our {\it ansatz}.
Terms of the form
\begin{equation}
\mbox{\rm Tr}\, 
\left[
  H_{\mu_1\cdots\mu_n\mu} p_\mu \gamma_5 
  \bar{\cal H}_{\mu_1\cdots\mu_n} \right]
\ , \qquad
\mbox{\rm Tr}\, 
\left[
  {\cal H}_{\mu_1\cdots\mu_n\mu} p_\mu \gamma_5 
  \bar{H}_{\mu_1\cdots\mu_n}
\right]
\end{equation}
can be shown to vanish by the heavy spin symmetry.
In the notation of Eq.~(\ref{Int for curlH}),
$d_{\rm S}=d_{{\rm S}0}$, $d_{\rm T}=d_{{\rm S}1}$ and 
$f_{\rm ST}=f_{{\rm S}0}$.
A new type of coupling present in Eq.~(\ref{general d f term})
also connects multiplets to others differing by $\Delta \ell=\pm2$.
These are the terms with odd (even) $n$ for $H$ (${\cal H}$)--type
fields. The interactions in Eq.~(\ref{general g term}) 
connecting multiplets
differing by $\Delta\ell=\pm1$ turn out not to contribute in our 
model. In the interest of simplicity we will consider all heavy 
mesons to have the same mass. This is clearly an approximation which 
may be improved in the future.

The rest frame {\it ans\"atze} for the bound state wave functions which
generalize Eq.~(\ref{curlH: matrix}) are (note $j_l=n+1/2$):
\begin{equation}
\left(\bar{H}_{i_1\cdots i_n}\right)_{\rm c} \rightarrow
\left\{
\begin{array}{l}
\displaystyle
\bar{h}^a_{i_1\cdots i_n,lh} \otimes
\left(
\begin{array}{cc}
0 & 0 \\ 1 & 0
\end{array}
\right)
\ , \qquad j_l=\ell+\frac{1}{2} \ ,
\\
\displaystyle
\bar{h}^a_{i_1\cdots i_n,lh} \otimes
\left(
\begin{array}{cc}
1 & 0 \\ 0 & 0
\end{array}
\right)
\ , \qquad j_l=\ell-\frac{1}{2}  \ ,
\end{array}
\right.
\label{general ansatz}
\end{equation}
with identical structures for $\bar{H} \rightarrow \bar{\cal H}$.
Note that again $a$, $l$, $h$ represent respectively the isospin, 
light spin and heavy spin bivalent indices.
Extracting a factor of 
$\hat{\mbox{\boldmath$x$}}\cdot\mbox{\boldmath$\tau$}$ as we did
before in Eqs.~(\ref{h: classical}) and (\ref{f: classical}) leads to
\begin{equation}
\bar{h}^a_{i_1\cdots i_n,lh} = 
\frac{u\left(\left\vert\mbox{\boldmath$x$}\right\vert\right)}%
{\sqrt{M}}
\left(\hat{\mbox{\boldmath$x$}}\cdot\mbox{\boldmath$\tau$}\right)_{ad}
\psi_{i_1\cdots i_n,dl}\left(k^\prime,k_3^\prime,r\right)\,\chi_h
\end{equation}
with similar notations.
The relevant wave--functions are the 
$\psi_{i_1\cdots i_n,dl}\left(k^\prime,k_3^\prime,r\right)$.
$k^\prime$ was defined in Eq.~(\ref{def: k prime});
we will see that it remains a good quantum number.
Since the terms which connect the positive parity ($H$ type) and
negative parity (${\cal H}$ type) heavy mesons 
(Eq.~(\ref{general g term})) vanish when the {\it ans\"atze}
(\ref{general ansatz}) are substituted, the baryon states associated 
with each type do not mix with each other in our model.
We thus list separately the potentials for each type.
For the $\ell=\mbox{even}$ baryons (associated with $H$ mesons),
\begin{eqnarray}
V\left[k'=0\right] &=&
- \frac{3}{2} d_{{\rm P}0} \, F'(0) \ ,
\nonumber\\
V\left[k'\neq0\right] &=& F'(0) \,
\left[ \begin{array}{cc}
\displaystyle
- \left(-1\right)^{k'} \frac{d_{{\rm P}(k'-1)}}{2} &
\displaystyle
-i\sqrt{\frac{2}{3}} f_{{\rm P}(k'-1)} \\
\displaystyle
 i\sqrt{\frac{2}{3}} f^{\ast}_{{\rm P}(k'-1)} &
\displaystyle
- \left(-1\right)^{k'} \frac{2k'+3}{2k'+1} \frac{d_{{\rm P}k'}}{2} 
\end{array} \right]
\ ,
\label{pot: gen p}
\end{eqnarray}
while for the $\ell=\mbox{odd}$ baryons (associated with ${\cal H}$
mesons), 
\begin{eqnarray}
V\left[k'=0\right] &=&
- \frac{3}{2} d_{{\rm S}0} \, F'(0) \ ,
\nonumber\\
V\left[k'\neq0\right] &=& F'(0) \,
\left[ \begin{array}{cc}
\displaystyle
- \left(-1\right)^{k'} \frac{d_{{\rm S}(k'-1)}}{2} &
\displaystyle
-i\sqrt{\frac{2}{3}} f_{{\rm S}(k'-1)} \\
\displaystyle
 i\sqrt{\frac{2}{3}} f^{\ast}_{{\rm S}(k'-1)} &
\displaystyle
- \left(-1\right)^{k'} \frac{2k'+3}{2k'+1} \frac{d_{{\rm S}k'}}{2} 
\end{array} \right]
\ .
\label{pot: gen s}
\end{eqnarray}
Details of the derivations of Eqs.~(\ref{pot: gen p}) and 
(\ref{pot: gen s}) are given in Appendix~\ref{app:a}.
The ordering of matrix elements in Eqs.~(\ref{pot: gen p}) and
(\ref{pot: gen s}), for a given $k'$, is such that the first heavy
meson has a light spin, $j_l = k' - \frac{1}{2}$ while the 
second has $j_l=k'+\frac{1}{2}$.  The $H$ type (${\cal H}$ type) 
channels with $k'=\mbox{even}$ (odd) involve two mesons with the 
same $\ell=k'$.  The $H$ type (${\cal H}$ type) channels with 
$k'=\mbox{odd}$ (even) involve two mesons differing by $\Delta
\ell=2$, {\it i.e.}, $\ell=k'-1$ and $\ell=k'+1$. This pattern is, for
convenience, illustrated in Table~\ref{table:4}.
\begin{table}[htbp]
\begin{center}
\begin{tabular}{cc|cc|cc}
& & \multicolumn{2}{c|}{$H$ mesons} & 
 \multicolumn{2}{c}{${\cal H}$ mesons} \\
$k'$ & $j_l$ & $\ell$ & \# & $\ell$ & \# \\
\hline\hline
$0$ & $1/2$ & $0$ & $1$ & $1$ & $1$ \\
\hline
$1$ 
 & \multicolumn{1}{c|}{$\begin{array}{c} $1/2$ \\ $3/2$ \end{array}$}
 & \multicolumn{1}{c}{$\begin{array}{c} $0$ \\ $2$ \end{array}$}
 & $0$ 
 & \multicolumn{1}{c}{$\begin{array}{c} $1$ \\ $1$ \end{array}$}
 & 1 \\
\hline
$2$ 
 & \multicolumn{1}{c|}{$\begin{array}{c} $3/2$ \\ $5/2$ \end{array}$}
 & \multicolumn{1}{c}{$\begin{array}{c} $2$ \\ $2$ \end{array}$}
 & $1$ 
 & \multicolumn{1}{c}{$\begin{array}{c} $1$ \\ $3$ \end{array}$}
 & 2 \\
\hline
$3$ 
 & \multicolumn{1}{c|}{$\begin{array}{c} $5/2$ \\ $7/2$ \end{array}$}
 & \multicolumn{1}{c}{$\begin{array}{c} $2$ \\ $4$ \end{array}$}
 & $0$ 
 & \multicolumn{1}{c}{$\begin{array}{c} $3$ \\ $3$ \end{array}$}
 & 1 \\
\end{tabular}
\end{center}
\caption[]{
Pattern of states for Eqs.~(\ref{pot: gen p}) and (\ref{pot: gen s}).
Note that $j_l=n+\frac{1}{2}$ is the light cloud spin of the heavy
meson.
The columns marked \# stand for the number of channels which are
expected to be bound, for that particular $k'$, according to the CQM.
}
\label{table:4}
\end{table}
Also shown, for each $k'$, are the number of channels which are
expected to be bound according to the CQM.

It is important to note that Table~\ref{table:4} holds for any value
of the angular momentum $r$, which is a good quantum number in our
model.  
For the reader's orientation, we now locate the previously considered
cases in Table~\ref{table:4}.  
The standard ``ground state'' heavy baryons discussed in 
section~\ref{sec:preliminaries} are made from the $H$ meson with
$\ell=0$ and $j_l=1/2$.
They have $r=0$ and $k'=0$.
The seven negative parity heavy baryons discussed in 
section~\ref{sec:preliminaries} also are made from the $H$ meson with
$\ell=0$ and $j_l=1/2$.
They still have $k'=0$, but now $r=1$.
The seven ``missing'' first excited heavy baryons discussed in 
section~\ref{sec:4} have $r=0$ and are made from the $\ell=1$, 
${\cal H}$ and ${\cal H}_\mu$ mesons with $j_l=1/2$ and 
$j_l=3/2$.
There should appear one bound state for $k'=0$, one bound state for
$k'=1$ and one bound state for $k'=2$ in the ``${\cal H}$--meson''
section of Table~\ref{table:4}.
Note that the number of states expected in the CQM model for $k'=2$ 
is listed in Table~\ref{table:4} as two, rather than one. In the 
absence of $\Delta \ell=2$ terms connecting ${\cal H}_\mu$ and
${\cal H}_{\mu\nu}$ (see the last term in 
Eq.~(\ref{general d f term})) $\ell$ would be conserved for our model 
and only the $\ell=1$ state would be relevant. This was the
approximation we made, for simplicity, in section~\ref{sec:4}. The
other entry would have $\ell=3$ and would decouple.  
When the $\Delta\ell=2$ mixing terms are turned on, the $\ell=1$ and
$\ell=3$, $k'=2$ 
channels will mix. One diagonal linear combination should be counted
against the $L_I=1$ CQM states and one against the $L_I=3$ CQM states.

To summarize: for the $H$--type mesons, the even $k'$ channels should
each have one bound state, while the odd $k'$ channels should have
none.  The situation is very different for the ${\cal H}$--type 
mesons; then the even $k'\neq0$ channels should contain two bound
states while the odd $k'$ channels should contain one bound state. The
$k'=0$ channel should have one bound state.

For the $H$--type meson case, the pattern of bound states mentioned
above would be achieved dynamically if the coupling constants 
satisfied:
\begin{eqnarray}
&& d_{{\rm P}0} > 0 \ ,
\nonumber\\
&& 
\left(-1\right)^{k'}
\left[
  d_{{\rm P}(k'-1)} d_{{\rm P}k'} \left( \frac{2k'+3}{2k'+1} \right)
  - \frac{8}{3} \left\vert f_{{\rm P}(k'-1)} \right\vert^2
\right]
< 0 \ , \quad (k'>0)
\nonumber\\
&&
d_{{\rm P}(k'-1)} + \left(\frac{2k'+3}{2k'+1} \right)
d_{{\rm P}k'} > 0
\ , \quad (k'=\mbox{odd}) \ .
\label{constraint: p}
\end{eqnarray}
These follow from requiring only one negative eigenvalue of 
Eq.~(\ref{pot: gen p}) for $k'=\mbox{even}$ and none for
$k'=\mbox{odd}$. Similarly requiring for the ${\cal H}$--type 
meson case in Eq.~(\ref{pot: gen s}), a negative eigenvalue for 
$k'=0$, one negative eigenvalue for $k'=\mbox{odd}$ and two 
negative eigenvalues for $k'>0$ and even leads to the criteria,
\begin{eqnarray}
&& d_{{\rm S}0} > 0 \ ,
\nonumber\\
&& 
\left(-1\right)^{k'}
\left[
  d_{{\rm S}(k'-1)} d_{{\rm S}k'} \left( \frac{2k'+3}{2k'+1} \right)
  - \frac{8}{3} \left\vert f_{{\rm S}(k'-1)} \right\vert^2
\right]
> 0 \ , \quad (k'>0)
\nonumber\\
&&
d_{{\rm S}(k'-1)} + \left(\frac{2k'+3}{2k'+1} \right)
d_{{\rm S}k'} > 0
\ , \quad (k'=\mbox{even}\neq0) \ .
\label{constraint: s}
\end{eqnarray}
{}From Eqs.~(\ref{constraint: p}) and (\ref{constraint: s}) it can be
seen that all the $d$'s are required to be positive.
Furthermore these equations imply that the 
$\left\vert f \right\vert$'s which connect heavy mesons with 
$\Delta \ell=2$ are relatively small (compared to the $d$'s) while the
$\left\vert f \right\vert$'s which connect heavy mesons with 
$\Delta \ell=0$ are relatively large. In detail this means that
$\left\vert f_{{\rm P}(k'-1)} \right\vert$ should be small for odd
$k'$ and large for even $k'$ with just the reverse for 
$\left\vert f_{{\rm S}(k'-1)} \right\vert$.
This result seems physically reasonable.

As in the example in the preceding section we should introduce the
collective variable $A(t)$ in order to define states of good isospin
and angular momentum. This again yields some splitting of the different 
$\left\vert \mbox{\boldmath$K$}^\prime + 
\mbox{\boldmath$J$}^{\rm sol} \right\vert$
members of each $k'$ bound state. Now, each $k'$ channel (except for 
$k'=0$) is described by a $2\times2$ matrix. Thus there will be an 
appropriate mixing angle $\theta$, analogous to the one introduced in 
Eq.~(\ref{rotation}), for each $k'$ and parity choice ({\it i.e.}, 
$H$--type or ${\cal H}$--type field). The collective Lagrangian is 
still given by Eq.~(\ref{def: Lcoll}) but, in the general case,
\begin{equation}
\chi_{\pm}(k') = \frac{1}{2k'(k'+1)}
\left[ 
  \frac{1}{2} \pm \left( k' + \frac{1}{2} \right) \cos 2\theta
\right]
\ .
\end{equation}
In this formula the different signs corresponds to the two possible
eigenvalues,
\begin{equation}
\lambda_{\pm} = 
\left[
  \frac{\left(-1\right)^{k'-1}}{4}
  \left( d_{(k'-1)} + \frac{2k'+3}{2k'+1} d_{k'} \right)
  \pm \frac{1}{4}
  \sqrt{ \left( d_{(k'-1)} - \frac{2k'+3}{2k'+1} d_{k'} \right)^2
    + \frac{32}{3} \left\vert f_{(k'-1)} \right\vert^2 
  }
\right]
F'(0)
\end{equation}
of the potential matrix.
For example, referring to Table~\ref{table:4}, we would expect the 
$k'=2$,
${\cal H}$--type meson case to provide two distinct bound states and
hence both $\chi_+(2,{\cal H})$ and $\chi_-(2,{\cal H})$ would be
non-zero. On the other hand, we would expect no bound states in 
the $k'=3$, $H$--type meson case so $\chi_{\pm}(3,H)$ should be 
interpreted as zero.

It is convenient to summarize the energies of the predicted states in
tabular form, generalizing the example presented in Table~\ref{tab:1}.
The situation for baryons with $\mbox{parity}=- (-1)^r$ 
(${\cal H}$--type mesons) is presented in Table~\ref{table:5}.
For definiteness we have made the assumption that the constraints
(\ref{constraint: s}) above are satisfied.
\begin{table}[htbp]
\begin{center}
\small
\begin{tabular}{c|c|c|c|c|c}
$I$ & $k'$ 
 & $\left\vert
      \mbox{\boldmath$K$}^\prime+\mbox{\boldmath$J$}^{\rm sol}
   \right\vert$
 & $V$ & $\alpha^2 \times H_{\rm coll}$
 & Candidates for $r=0$ 
\\
$=J^{\rm sol}$ & & & & & missing states ~~~~~~
\\
\hline
 & $2n-1$ & $2n-1$ & $\lambda_+$ & $n(2n-1)\chi_-^2$
 & $\left\{
     \Lambda\left((2n-3/2)^-\right)\,,\,
     \Lambda\left((2n-1/2)^-\right)
   \right\}$
\\
\cline{2-6}
{\large$0$} & $2n$ & $2n$ & $\lambda_+$ & $n(2n+1)\chi_+^2$
 & $\left\{
     \Lambda\left((2n-1/2)^-\right)\,,\,
     \Lambda\left((2n+1/2)^-\right)
   \right\}$
\\
 & & & $\lambda_-$ & $n(2n+1)\chi_-^2$ & $''$ 
\\
\hline
 & $2n-1$ & $2n-2$ & & $n(2n-1)\chi_+^2+1-2n\chi_+$ & 
\\
 & & $2n-1$ & $\lambda_+$ 
 & $n(2n-1)\chi_+^2+1-\chi_+$ 
 & $\left\{
     \Sigma\left((2n-3/2)^-\right)\,,\,
     \Sigma\left((2n-1/2)^-\right)
   \right\}_1$
\\
 & & $2n$ & & $n(2n-1)\chi_+^2+1+(2n-1)\chi_+$ & 
\\
\cline{2-6}
 & $2n$ & $2n-1$ & & $n(2n+1)\chi_+^2+1-(2n+1)\chi_+$
 & $\left\{
     \Sigma\left((2n-3/2)^-\right)\,,\,
     \Sigma\left((2n-1/2)^-\right)
   \right\}_2$
\\
 {\large$1$} & & $2n$ & $\lambda_+$ & $n(2n+1)\chi_+^2+1-\chi_+$ & 
\\
 & & $2n+1$ & & $n(2n+1)\chi_+^2+1+2n\chi_+$
 & $\left\{
     \Sigma\left((2n+1/2)^-\right)\,,\,
     \Sigma\left((2n+3/2)^-\right)
   \right\}_3$
\\
\cline{3-6}
 & & $2n-1$ & & $n(2n+1)\chi_-^2+1-(2n+1)\chi_-$
 & $\left\{
     \Sigma\left((2n-3/2)^-\right)\,,\,
     \Sigma\left((2n-1/2)^-\right)
   \right\}_4$
\\
 & & $2n$ & $\lambda_-$ & $n(2n+1)\chi_-^2+1-\chi_-$ & 
\\
 & & $2n+1$ & & $n(2n+1)\chi_-^2+1+2n\chi_-$
 & $\left\{
     \Sigma\left((2n+1/2)^-\right)\,,\,
     \Sigma\left((2n+3/2)^-\right)
   \right\}_5$
\\
\end{tabular}
\end{center}
\caption[]{
Contributions to energies of the new predicted states made from 
${\cal H}$--type heavy mesons.
Note that $n$ is a positive integer.
The $n=0$ case is given in Table~\ref{tab:1}.
The $\lambda_+$ entries in the $V$ column are more tightly bound than
the $\lambda_-$ entries.
$\left\vert \mbox{\boldmath$K$}^\prime+\mbox{\boldmath$J$}^{\rm sol}
\right\vert$ is the light part of the heavy baryon angular momentum
for $r=0$ (See Eq.~(\ref{baryon jlight}).).
}
\label{table:5}
\end{table}
In order to explain Table~\ref{table:5} let us ask which states
correspond to the ($L_I=3$, $L_E=0$) states in the CQM.
Reference to Table~\ref{table:1} shows that three negative parity
$\Lambda$--type heavy multiplets and one negative parity
$\Sigma$--type heavy multiplet should be present.
The correspondence in Eq.~(\ref{correspondence:2}) instructs us to set
$r=0$ and, noting Eq.~(\ref{def: k prime}) , to identify
\begin{equation}
\mbox{\boldmath$K$}^\prime + \mbox{\boldmath$J$}^{\rm sol}
\leftrightarrow 
\mbox{\boldmath$L$}_I + \mbox{\boldmath$S$}
\ .
\end{equation}
The $\Lambda$--type particles are of type c) in 
Eq.~(\ref{condition:CQM}) so we must take $S=1$.
Hence, since $J^{\rm sol}=0$ for $\Lambda$--type particles, we learn
that $k'$ can take on the values $2$, $3$ and $4$.
For $k'=2$, the second line of the $k'$ column yields two possible
multiplets (energies $\lambda_+$ and $\lambda_-$) with $n=1$ and
structure $\left\{ \Lambda\left(\frac{3}{2}^-\right)\,,\,
\Lambda\left(\frac{5}{2}^-\right)  \right\}$.
We should choose one of these to be associated with ($L_I=3$, $L_E=0$)
and the other with ($L_I=1$, $L_E=0$) in the CQM.
We remind the reader that $\ell$ is not a good quantum number so that 
the correspondence $\mbox{\boldmath$\ell$} \leftrightarrow
\mbox{\boldmath$L$}_I$ in Eq.~(\ref{correspondence:2}) only holds when
the $\Delta \ell=2$ mixing terms are neglected.
For $k'=3$, the first line of the $k'$ column correctly yields one
multiplet with $n=2$ and structure 
$\left\{ \Lambda\left(\frac{5}{2}^-\right)\,,\,
\Lambda\left(\frac{7}{2}^-\right)  \right\}$.
For $k'=4$, the second line of the $k'$ column yields two multiplets
with $n=2$ and structure
$\left\{ \Lambda\left(\frac{7}{2}^-\right)\,,\,
\Lambda\left(\frac{9}{2}^-\right)  \right\}$.
One of these is to be associated with ($L_I=3$, $L_E=0$) and the other
with ($L_I=5$, $L_E=0$) in the CQM.
Now let us go on to the $\Sigma$--type heavy multiplets.
These are of type d) in Eq.~(\ref{condition:CQM}) and yield $S=0$.
Hence $\mbox{\boldmath$K$}^\prime + \mbox{\boldmath$J$}^{\rm sol}
\leftrightarrow \mbox{\boldmath$L$}_I$ and 
$\left\vert\mbox{\boldmath$K$}^\prime + 
\mbox{\boldmath$J$}^{\rm sol} \right\vert=3$.
Five candidates for this 
$\left\{ \Sigma\left(\frac{5}{2}^-\right)\,,\,
\Sigma\left(\frac{7}{2}^-\right)  \right\}$
multiplet are shown in the last column of Table~\ref{table:5}.
These consecutively correspond to the choices $n=2$, $2$, $1$, $2$,
$1$ in the $\left\vert\mbox{\boldmath$K$}^\prime + 
\mbox{\boldmath$J$}^{\rm sol} \right\vert$ column.
As before it is necessary for an exact correspondence with the CQM
that one of these should be dynamically favored (much more tightly
bound) over the others.
Again, note that the choice
$\left\vert\mbox{\boldmath$K$}^\prime + 
\mbox{\boldmath$J$}^{\rm sol} \right\vert=3$ does not uniquely
constrain the value of $\ell$.

Next, the situation for baryons with
$\mbox{parity}=\left(-1\right)^{r}$ ($H$--type baryons) is presented
in Table~\ref{table:6}..
\begin{table}[htbp]
\begin{center}
\small
\begin{tabular}{c|c|c|c|c|c}
$I$ & $k'$ 
 & $\left\vert
      \mbox{\boldmath$K$}^\prime+\mbox{\boldmath$J$}^{\rm sol}
   \right\vert$
 & $V$ & $\alpha^2 \times H_{\rm coll}$
 & Candidates for $r=0$ 
\\
$=J^{\rm sol}$ & & & & & missing states ~~~~~~
\\
\hline
$0$ & $2n$ & $2n$ & $\lambda_+$ & $n(2n+1)\chi_+^2$
 & $\left\{
     \Lambda\left((2n-1/2)^+\right)\,,\,
     \Lambda\left((2n+1/2)^+\right)
   \right\}$
\\
\hline
 & & $2n-1$ & & $n(2n-1)\chi_+^2+1-(2n+1)\chi_+$ 
 & $\left\{
     \Sigma\left((2n-3/2)^+\right)\,,\,
     \Sigma\left((2n-1/2)^+\right)
   \right\}_1$
\\
 1 & $2n$ & $2n$ & $\lambda_+$ & $n(2n+1)\chi_+^2+1-\chi_+$
 & $\left\{
     \Sigma\left((2n-1/2)^+\right)\,,\,
     \Sigma\left((2n+1/2)^+\right)
   \right\}_2$
\\
 & & $2n+1$ & & $n(2n+1)\chi_+^2+1+2n\chi_+$
 & $\left\{
     \Sigma\left((2n+1/2)^+\right)\,,\,
     \Sigma\left((2n+3/2)^+\right)
   \right\}_3$
\\
\end{tabular}
\end{center}
\caption[]{
Contributions to energies of the new predicted states made from 
$H$--type heavy mesons.
Other details as for Table~\ref{table:5}.
}
\label{table:6}
\end{table}
For definiteness we have made the assumption that the constraints
(\ref{constraint: p}) above are satisfied.
This eliminates the odd $k'$ states and agrees with the CQM counting.
For example, we ask which states correspond to the ($L_I=2$, $L_E=0$)
states in the CQM.
Reference to Table~\ref{table:1} shows that one positive parity
$\Lambda$--type heavy multiplet and three positive parity
$\Sigma$--type heavy multiplets should be present.
For $r=0$ we have the correspondence $\mbox{\boldmath$K$}^\prime +
\mbox{\boldmath$J$}^{\rm sol} \leftrightarrow \mbox{\boldmath$L$}_I +
\mbox{\boldmath$S$}$.
The $\Lambda$--type particles are of type a) in
Eq.~(\ref{condition:CQM}) so we must set $k'=2$.
The first line in Table~\ref{table:6} then yields, with $n=1$ the
desired $\left\{\Lambda\left(\frac{3}{2}^+\right) \,,\,
\Lambda\left(\frac{5}{2}^+\right)\right\}$ heavy multiplet.
The $\Sigma$ particles are of type b) in Eq.~(\ref{condition:CQM}) so
that $\left\vert \mbox{\boldmath$K$}^\prime + 
\mbox{\boldmath$J$}^{\rm sol} \right\vert$ can take on the values $1$,
$2$ and $3$.
The last three lines in Table~\ref{table:6}, with $n=1$, give the
desired multiplets:
$\left\{\Sigma\left(\frac{1}{2}^+\right) \,,\,
\Sigma\left(\frac{3}{2}^+\right)\right\}$,
$\left\{\Sigma\left(\frac{3}{2}^+\right) \,,\,
\Sigma\left(\frac{5}{2}^+\right)\right\}$ and 
$\left\{\Sigma\left(\frac{5}{2}^+\right) \,,\,
\Sigma\left(\frac{7}{2}^+\right)\right\}$.
In this case all the states should be bound so that the splittings 
due to $H_{\rm coll}$ are desired to be relatively small.
The present structure is simpler than the one shown in
Table~\ref{table:5} for the ${\cal H}$--type cases.

\section{Discussion}
\label{sec:6}

In this paper, we have pointed out the problem of getting, in the
framework of a bound state picture, the excited states which are
expected on geometrical grounds from the constituent quark model.
We treated the heavy baryons and made use of the Isgur--Wise heavy
spin symmetry.
The approach may also provide some insight into the understanding of
light excited baryons.
The key problem to be solved is the introduction of an additional
``source'' of angular momentum in the model.
It was noted that this might be achieved in a simple way by
postulating that excited heavy mesons, which have ``locked--in" 
angular momentum, are bound in the background Skyrmion field.
The model was seen to naturally have the correct kinematical structure
in order to provide the excited states which were missing in earlier
models.

An important aspect of this work is the investigation of 
which states in the model are actually bound.
This is a complicated issue since there are many interaction terms
present with {\it a priori} unknown coupling constants.
Hence, for the purpose of our initial investigation we included only
terms with the minimal interactions of the light pseudoscalar mesons.
The large $M$ limit was also assumed and nucleon recoil as well as
mass splittings among the heavy excited meson multiplets were
neglected. We expect, based on previous work, that the most important 
improvement of the present calculation would be to include the 
interactions of the light vector mesons. It is natural to expect 
that possible interactions of the light higher spin mesons 
also play a role.  In the calculation of the ground state heavy baryons 
the light vectors were actually slightly more important than the light 
pseudoscalars and reinforced the binding due to the latter.
Another complicating factor is the presence, expected from
phenomenology, of radially excited mesons along with orbitally excited
ones.

It is interesting to estimate which of the first excited states,
discussed in section~\ref{sec:4}, are bound.
The criteria for actually obtaining the missing states in the model
with only light pseudoscalars present are given in 
Eqs.~(\ref{dst constraints}) and (\ref{fst constraints}).
Based on the use of chiral symmetry for relating the coupling constants
to axial matrix elements and using a quark model argument to estimate
the axial matrix elements, Falk and Luke~\cite{Falk-Luke}
presented the estimates (their Eqs.~(2.23) and (2.24))
$d_{\rm T} = 3 d_{\rm S} = d$ and 
$\left\vert f_{\rm ST} \right\vert = \frac{2}{\sqrt{3}} d$.
With these estimates 
Eqs.~(\ref{dst constraints}) and (\ref{fst constraints}) are
satisfied.
Note that $d>0$ provides binding for the ground state heavy baryons.
However we have checked this and find that, although we are
in agreement for $\left\vert f_{\rm ST} \right\vert$ we obtain instead 
$d_{\rm T} = 3 d_{\rm S} = - d$.
Assuming that this is the case then it is easy to see that the 
only bound multiplet will have $k'=1$.
This leads to the desired $\Sigma$--type multiplet and one of the
three desired $\Lambda$--type multiplets being bound, but not the
$k'=0$ and $2$, $\Lambda$--type multiplets.
Clearly, it is important to make a more detailed calculation of the
light meson--excited heavy meson coupling constants.
We also plan to investigate the effects of including light vector
mesons in the present model.
It is hoped that the study of these questions will lead to a better
understanding of the dynamics of the excited heavy particles.

Finally we would like to add a few remarks on studies 
of the excited ``light'' hyperons within 
the bound state approach to the SU(3) Skyrme model. 
In that model the heavy spin symmetry is not maintained since 
the vector counterpart of the kaon, the $K^\ast$, is omitted;
while the kaons themselves couple to the pions as prescribed by 
chiral symmetry. On the other hand the higher 
orbital angular momentum channels ({\it i.e.} $r\ge2$) have 
been extensively studied. The first study 
was performed by the SLAC group \cite{Kr86}.
However, they were mostly interested in the amplitudes for 
kaon--nucleon scattering and for simplicity omitted 
flavor symmetry breaking terms in the effective Lagrangian.
Hence they did not find any bound states, except for 
zero modes. These symmetry breaking terms were, however, 
included in the scattering analysis of all higher orbital angular 
momentum channels by Scoccola \cite{Sc90}. The only bound states 
he observed were those for P-- and S--waves. After collective 
quantization these are associated with the ordinary hyperons 
and the $\Lambda(1405)$. As a matter of fact these states 
were already found in the original study by Callan and Klebanov 
\cite{Callan-Klebanov}. 
It is clear that the orbital excitations found in the bound 
state approach to the Skyrme model should be identified as the 
$\ell=0$ states. Furthermore when the dynamical coupling of 
the collective coordinates ($A,\mbox{\boldmath $\Omega$}$) 
is included in the scattering analysis \cite{Sch92} the only 
resonances which are observed obey the selection rule 
 $|J-1/2|\le r \le |J+1/2|$, where $r$ denotes the kaon 
orbital angular momentum. This rule is consistent 
with $\ell=0$ in our model. In order to find states with 
$\ell\ne0$ in this model one would also have to include
pion fluctuations besides the kaon fluctuations for the 
projectile--state. As indicated in section III these fluctuating 
fields should be coupled to carry the good quantum number $\ell$.
The full calculation would not only require this complicated 
coupling but also an expansion of the Lagrangian up to 
fourth order in the meson fluctuations off the background 
soliton. Such a calculation seems impractical, 
indicating that something like our present approximation, 
which treats these 
coupled states as elementary particles, is needed.

\acknowledgments

We would like to thank Asif Qamar for helpful discussions.
This work has been supported in part by the US DOE under contract
DE-FG-02-85ER 40231 and by the DFG under contract
Re 856/2--3.

\appendix

\section{Classical Potential}
\label{app:a}
Here we will show how to compute the relevant matrix 
elements associated with the classical potential.

For any fixed value of  $k^{\prime}\ne 0$ the heavy meson 
light cloud spin (${\mbox{\boldmath$J$}}_{\rm light}$) 
takes the values $j_{l}=k^{\prime}\mp \frac{1}{2}$ 
since 
${\mbox{\boldmath$K$}}^{\prime}= {\mbox{\boldmath$J$}}_{\rm light} + 
{\mbox{\boldmath$I$}}_{\rm light}$
, where 
$\mbox{\boldmath$I$}_{\rm light}$ is the heavy meson isospin. Hence
the classical potential will be, in general, a $2\times 2$ matrix
schematically  
represented as 
\begin{equation}
V\left(k^{\prime}\ne 0\right)=
\left( \begin{array}{cc}
\langle 
 H_{\mu_1\cdots\mu_{k^\prime-1}}|V|H_{\mu_1\cdots\mu_{k^\prime-1}}
\rangle 
& 
\langle 
 H_{\mu_1\cdots\mu_{k^\prime-1}} |V|H_{\mu_1\cdots\mu_{k^\prime}}
\rangle
\\ 
\langle 
 H_{\mu_1\cdots\mu_{k^\prime~~}} |V| H_{\mu_1\cdots\mu_{k^\prime-1}} 
\rangle  
&  
\langle
 H_ {\mu_1\cdots\mu_{k^{\prime~~}}}|V| {H}_{\mu_1\cdots\mu_{k^\prime}}
\rangle
\end{array}\right)
 \ .
\label{kprime matrix}
\end{equation}
Here $|H_{\mu_1\cdots\mu_{k^{\prime}-1}}\rangle$ corresponds to the  
$j_{l}=k^{\prime}-\frac{1}{2}$  state 
while 
$|H_{\mu_1\cdots\mu_{k^{\prime}}}\rangle$ corresponds to 
 $j_{l}=k^{\prime}+\frac{1}{2}$.
 In order to compute the potential 
there is no need to distinguish even parity heavy mesons ${\cal H}$ 
from odd parity ones $H$. 
The diagonal matrix elements are 
obtained by substituting the appropriate rest frame {\it ansatz}
(\ref{general ansatz}) into the general potential term as:
\begin{eqnarray}
&-& i M\, d_{n}\, (-1)^n \int d^3x \, 
\mbox{Tr}\, \left[ H_{\mu_1\cdots\mu_{n}}\gamma_{\alpha}
\gamma_5 p_{\alpha}\bar{H}_{\mu_1\cdots\mu_{n}}\right] \nonumber \\ 
&=& d_n\, \frac{F^{\prime}(0)}{2}\,(-1)^n  \int d\Omega \, 
\psi^{\ast}_{i_{1}\cdots i_{n},dl}
\left(k^{\prime},k^{\prime}_3,r\right) 
{\mbox{\boldmath$\sigma$}}_{ll^{\prime}}\cdot
{\mbox{\boldmath$\tau$}}_{dd^{\prime}}
\psi_{i_{1}\cdots i_{n},d^{\prime}l^{\prime}}
\left(k^{\prime},k^{\prime}_3,r\right) \ ,
\label{diagonal element}
\end{eqnarray}
where $j_{l}=n + \frac{1}{2}$ and $n=k^{\prime}\mp 1$ for the two 
diagonal matrix elements. 
The operator which mesures the total 
light cloud  spin $j_{l}$  is 
\begin{eqnarray}
 \left(J^a_{\rm light}\right)_{i_1 j_1,\cdots,i_n j_n;ll^{\prime}} 
&=&\frac{\sigma^a_{ll^{\prime}}}{2}
\otimes\delta_{i_1j_1}\otimes\cdots\otimes\delta_{i_nj_n}+
\delta_{ll^{\prime}}\otimes\left(-i\epsilon_{ai_1j_1}\right)\otimes
\delta_{i_2j_2}\otimes\cdots\otimes\delta_{i_nj_n}  \nonumber \\
&+&
\cdots +\delta_{ll^{\prime}}\otimes\delta_{i_1j_1}\otimes\cdots\otimes
\delta_{i_{n-1}j_{n-1}}\otimes\left(-i\epsilon_{ai_n j_n}\right) \ .
\label{jlight}
\end{eqnarray}
where $\epsilon_{aij}$ is the totally antisymmetric 
tensor. 
The isospin operator is
\begin{equation}
\mbox{\boldmath$I$}_{\rm light} = 
\frac{\mbox{\boldmath$\tau$}}{2} \ .
\end{equation}
We can write Eq.~(\ref{jlight}) compactly in the following way 
\begin{equation}
{\mbox{\boldmath$J$}}_{\rm light}
=\mbox{\boldmath$s$}+\hat{\mbox{\boldmath$l$}} \ ,
\end{equation}
where $\mbox{\boldmath$s$}\equiv\frac{\mbox{\boldmath$\sigma$}}{2}$.
Due to the 
total symmetrization of the vectorial indices we have $\hat{l}=n$.
We want to stress that $\mbox{\boldmath$s$}$ and 
$\hat{\mbox{\boldmath$l$}}$ do not necessarily agree with 
$\mbox{\boldmath$S$}_{\rm light}$ and $\mbox{\boldmath$\ell$}$.
Indeed for $\Phi_{ld}$ associated with ${\cal H}$ in 
Eq.~(\ref{f: classical}), $\hat{l}=0$ and 
$\mbox{\boldmath$J$}_{\rm light}=\mbox{\boldmath$s$}=
\mbox{\boldmath$S$}_{light} + \mbox{\boldmath$\ell$}$ while
for associated $\Phi_{i,ld}$ with ${\cal H}_\mu$, $\hat{l}=1$.
Now we have,
for fixed $n=j_l-\frac{1}{2}$, the following 
useful result:
\begin{equation}
\int d\Omega \, \psi^{\ast}\mbox{\boldmath$s$}\psi =  
\frac{\int d\Omega \psi^{\ast}\,
  \left(\mbox{\boldmath$s$}\cdot\mbox{\boldmath$J$}_{\rm light}\right)
\psi}{j_l (j_l+1)} \int d\Omega \, 
\psi^{\ast}{\mbox{\boldmath$J$}}_{\rm light}\psi 
=
\frac{1}{2\, j_{l}} \int d\Omega \, 
\psi^{\ast}{\mbox{\boldmath$J$}}_{\rm light}\psi \ .
\label{wigner}
\end{equation}
By using Eq.~(\ref{wigner}) we can 
write Eq.~(\ref{diagonal element}) as
\begin{eqnarray}
&~&(-1)^n\,  d_n \frac{F^{\prime}(0)}{j_{l}} 
\int d\Omega \, 
\psi^{\ast}\left(k^{\prime},k^{\prime}_3,r\right)
{\mbox{\boldmath$J$}}_{\rm light}\cdot 
{\mbox{\boldmath$I$}}_{\rm light} \psi 
\left(k^{\prime},k^{\prime}_3,r\right)\nonumber \\  
&=&
(-1)^n\, d_n \frac{F^{\prime}(0)}{2 j_{l}} 
\left[k^{\prime}(k^{\prime}+1) - j_{l}(j_{l}+1) - \frac{3}{4} \right] 
\ .
\label{second step}
\end{eqnarray}
For $j_l=k^{\prime}\mp \frac{1}{2}$ we get the diagonal matrix
elements for both, the $H$ type as well as the ${\cal H}$ type fields
\begin{equation}
(-1)^{k^\prime-1}\frac{F^\prime(0)}{2}\cdot
\left\{\begin{array}{ll}
\displaystyle
d_{k^\prime-1} \ , & j_l=k^\prime-\frac{1}{2} \ ,\\
\displaystyle
d_{k^\prime}\, \left(\frac{2\,k^{\prime}+3}{2\, k^\prime+1}\right) \ ,
& j_l=k^{\prime}+\frac{1}{2} \ ,
\end{array}\right.
\label{def: final diag elements}
\end{equation}
where we used $n=j_l-1/2$.

{}For the non--diagonal matrix elements we consider the contribution 
to the potential due to the following 
$f$ type term:
\begin{eqnarray}
&& -  i M\, f_{n}\, (-1)^n \int d^3x \, 
\mbox{Tr}\, \left[ H_{\mu_1\cdots\mu_{n}}p_{\mu} 
\gamma_5 \bar{H}_{\mu_1\cdots\mu_{n}\mu}\right] \nonumber \\ 
&=& i f_n\, \frac{F^{\prime}(0)}{2}\,  \int d\Omega \, 
\psi^{\ast}_{i_{1}\cdots i_{n},dl}
\left(k^{\prime},k^{\prime}_3,r\right) 
\tau^i_{dd^{\prime}}
\psi_{i_{1}\cdots i_{n}i,d^{\prime}l}
\left(k^{\prime},k^{\prime}_3,r\right) \ .
\label{non diagonal element}
\end{eqnarray}
This corresponds to the transition between $j_l=n+\frac{1}{2}$ and 
$j_l=n+\frac{3}{2}$ states.
Now we notice that by construction any wave function $\psi$ must 
satisfy the condition
\begin{equation}
\left(P^{3/2}\right)_{ii_{1};ll^{\prime}}
\psi_{i_1i_2\cdots i_n,dl^{\prime}}=
\psi_{ii_2\cdots i_n,dl} \ ,
\label{rarita2}
\end{equation}
where $P^{3/2}$ is the spin $3/2$ projection operator 
\begin{equation}
\left(P^{3/2}\right)_{ik;ll^{\prime}}=\frac{2}{3} 
\left(\delta_{ik}\delta_{ll^{\prime}}- \frac{i}{2}\epsilon_{jik} 
\sigma^j_{ll^{\prime}}\right) \ .
\end{equation}
The condition (\ref{rarita2}) yields the following identity
\begin{eqnarray}
 &\int& d\Omega \, 
\psi^{\ast}_{i_{1}\cdots i_{n},dl}
\left(k^{\prime},k^{\prime}_3,r\right) 
\tau^i_{dd^{\prime}}
\psi_{i_{1}\cdots i_{n}i,d^{\prime}l}
\left(k^{\prime},k^{\prime}_3,r\right)= \nonumber \\
 &\int& d\Omega \, 
\psi^{\ast}_{i_{1}\cdots i_{n},dl}
\left(k^{\prime},k^{\prime}_3,r\right) 
\tau^j_{dd^{\prime}}\left(P^{3/2}\right)_{jk;ll^{\prime}}
\psi_{i_{1}\cdots i_{n}k,d^{\prime}l^{\prime}}
\left(k^{\prime},k^{\prime}_3,r\right) \ . 
\end{eqnarray}
Using the fact that $P^{3/2}\mbox{\boldmath$\tau$}$ commutes
with $\mbox{\boldmath$K$}^\prime$, we get
\begin{equation}
\left(P^{3/2}\right)_{jk;ll^{\prime}}\tau^{k}_{dd^{\prime}}
\psi_{i_{1}\cdots i_{n},dl^{\prime}}
\left(k^{\prime},k^{\prime}_3,r\right) 
= N \psi_{i_{1}\cdots i_{n}j,dl}
\left(k^{\prime},k^{\prime}_3,r\right) \ , 
\end{equation}
where $N$ is a normalization constant.
It is evaluated as
\begin{equation}
|N|^2=\int d\,\Omega\, 
\psi^{\ast}_{i_{1}\cdots i_{n},dl}
\left(k^{\prime},k^{\prime}_3,r\right) 
\tau^{c}_{dd^{\prime}}
\left(P^{3/2}\right)_{ck;ll^{\prime}}
\tau^{k}_{d^{\prime}d^{\prime\prime}}
\psi_{i_{1}\cdots i_{n},d^{\prime\prime}l^{\prime}}
\left(k^{\prime},k^{\prime}_3,r\right)=\frac{8}{3} \ .
\end{equation}
The non--diagonal matrix element is, up to a phase factor  
 in Eq.~(\ref{non diagonal element})  
\begin{equation}
i f_n \, F^{\prime}(0) \sqrt{\frac{2}{3}} \ , \quad ~~~\forall 
\, k^{\prime}\neq 0\ .
\end{equation}

{}For $k^{\prime}=0$ we have only one diagonal element with 
$j_{l}=\frac{1}{2}$. The second line of 
Eq.~(\ref{def: final diag elements})  
provides
\begin{equation}
V(k^{\prime}=0)=-\frac{3}{2}F^{\prime}(0)d_0\ .
\end{equation}

\section{Collective Lagrangian}
\label{app:b}

Here the relevant 
matrix elements associated with the collective coordinate Lagrangian 
are computed. 
We will restrict $k'$ to be nonzero since
there is no contribution 
for $k^{\prime}=0$ 
to the collective Lagrangian.

The kinetic Lagrangian for $H$ type and ${\cal H}$ type fields is:
\begin{equation}
{\cal L}_{\rm kin}=+i\,M V_{\mu}\sum_{n} (-1)^n \mbox{Tr}\, 
\left[ H_{\mu_1\cdots\mu_n} D_\mu \bar{H}_{\mu_1\cdots\mu_n} \right]
-i\,M V_{\mu}\sum_n (-1)^n 
\mbox{Tr}\, 
\left[ {\cal H}_{\mu_1\cdots\mu_n} D_\mu 
\bar{{\cal H}}_{\mu_1\cdots\mu_n} \right]
 \ .
\end{equation}
In the following we will not distinguish between the 
$H$ and ${\cal H}$ types of field.  
We need to consider the collective coordinate Lagrangian  for  a given 
 $k^{\prime}$ classical bound channel 
in the heavy meson rest frame. 
{}For $k^{\prime}\neq 0 $  the bound state wave--function can 
schematically be represented as
\begin{equation}
|Bound~State; k^{\prime} \rangle
= \alpha\, |H_{\mu_1\cdots\mu_{k^{\prime}-1}} \rangle
+ \beta\, |H_{\mu_1\cdots\mu_{k^{\prime}}} \rangle \ ,
\end{equation}
where $|\alpha|^2 + |\beta|^2=1$.

The collective coordinate Lagrangian 
($\delta L_{\rm coll}$), induced by the heavy meson kinetic term, 
is obtained by   
generalizing 
Eqs.~(\ref{def: collective}) and (\ref{def: curl collective}) 
to the higher excited heavy  
meson fields, introducing the collective coordinate $A(t)$ rotation
via  
\begin{equation}
\bar{H}_{i_1\cdots i_n}(\mbox{\boldmath$x$},t)=
A(t)\bar{H}_{i_1\cdots i_n {\rm c}}
(\mbox{\boldmath$x$})\ ,
\end{equation}
where the $\bar{H}_{i_1\cdots i_n {\rm c}}
(\mbox{\boldmath$x$})$ classical 
ansatz is given in Eq.~(\ref{general ansatz}). 
The contribution for fixed $k^{\prime}\neq 0$ is:
\begin{eqnarray}
\delta L_{\rm coll}&=& -\Omega^{q} \, 
\left[|\alpha|^2 \int d\,\Omega\, 
\psi^{\ast}_{i_{1}\cdots i_{k^{\prime}-1},dl}
\left(k^{\prime},k^{\prime}_3,r\right) 
\frac{\tau^q_{dd^{\prime}}}{2}
\psi_{i_{1}\cdots i_{k^{\prime}-1},d^{\prime}l} 
\left(k^{\prime},k^{\prime}_3,r\right) \right.\nonumber \\ 
&~&+\left.|\beta|^2 \int d\,\Omega \, 
\psi^{\ast}_{i_{1}\cdots i_{k^{\prime}},dl}
\left(k^{\prime},k^{\prime}_3,r\right) 
\frac{\tau^q_{dd^{\prime}}}{2}
\psi_{i_{1}\cdots i_{k^{\prime}},d^{\prime}l}
\left(k^{\prime},k^{\prime}_3,r\right) \right]\nonumber \\
&\equiv&-|\alpha|^2 
\int d\Omega\,
\psi^{\ast}
\left(k^{\prime},k^{\prime}_3,j_l=k^{\prime}-1/2\right)
\mbox{\boldmath$\Omega$}\cdot\mbox{\boldmath$I$}_{\rm light}
\psi
\left(k^{\prime},k^{\prime}_3,j_l=k^{\prime}-1/2\right)
\nonumber \\
&~&-|\beta|^2 
\int d\Omega
\psi^{\ast}
\left(k^{\prime},k^{\prime}_3,j_l=k^{\prime}+1/2\right)
\mbox{\boldmath$\Omega$}\cdot\mbox{\boldmath$I$}_{\rm light}
\psi
\left(k^{\prime},k^{\prime}_3,j_l=k^{\prime}+1/2\right)
\ ,
\label{collective kinetic term}
\end{eqnarray}
where the over all minus sign in Eq.~(\ref{collective kinetic term})
is required, as explained in section \ref{sec:4}.
According to the Wigner-Eckart theorem:
\begin{equation}
\int d \Omega \, \psi^{\ast} 
\mbox{\boldmath$I$}_{\rm light}
\psi = 
\frac{\left[k^{\prime}(k^{\prime}+1)-j_l(j_l+1) + \frac {3}{4}\right]}
{2\,k^{\prime}(k^{\prime}+1)} 
\int d\Omega \, \psi^{\ast}
\mbox{\boldmath$K$}^{\prime}
\psi
\ ,
\label{collective  matrix element}
\end{equation}
we thus obtain the following heavy meson contribution to the
collective coordinate Lagrangian for 
$k^{\prime}\neq 0$
\begin{equation}
\delta L_{\rm coll}
= 
-\chi(k^{\prime})\,
\mbox{\boldmath$\Omega$}\cdot\mbox{\boldmath$K$}^{\prime}
\ .
\end{equation}
The quantity $\chi(k')$ is given by
\begin{equation}
 \chi(k^{\prime})=\frac{1}{2\,k^{\prime}(k^{\prime}+1)}
\left[ \frac{1}{2} \pm \left(k^{\prime} + \frac{1}{2}\right) 
\cos 2 \theta\right] \ ,
\label{collective chi}
\end{equation}
where $|\alpha|^2 - |\beta|^2=\pm \cos 2 \theta$ was used.
In Eq.~(\ref{collective chi}) 
the $\pm$ sign corresponds to the two possible eigenvalues in the 
potential matrix for given $k^{\prime}\neq 0$.

\end{document}